%% file: QND_paper.tex
\documentclass[aps,pra,twocolumn,superscriptaddress,floatfix]{revtex4-1}
\usepackage{graphicx}
\usepackage{dcolumn}
\usepackage{bm}
\usepackage{amssymb}
\usepackage{amsmath}
\usepackage{color}
\usepackage{here}
\usepackage{braket}
\graphicspath{{./figure/},{./s_figure/}}

\usepackage{lipsum}
\usepackage{xcite}
\externalcitedocument[A-]{QND_paper}
\usepackage{xr}
\begin{document}

\title{Quantum Nondemolition Gate Operations and Measurements\\ in Real Time on Fluctuating Signals}

\input author_list.tex

\begin{abstract}
We demonstrate an optical quantum nondemolition (QND) interaction gate with a bandwidth of about 100~MHz.
Employing this gate, we are able to perform QND measurements in real time on randomly fluctuating signals.
Our QND gate relies upon linear optics and offline-prepared squeezed states.
In contrast to previous demonstrations on narrow sideband modes, our gate is compatible with non-Gaussian quantum states temporally localized in a wave-packet mode, 
and thus opens the way for universal gate operations and realization of quantum error correction.
\end{abstract}


\maketitle

{\it Introduction}.---A quantum nondemolition (QND) interaction enables indirect and nondestructive measurements of quantum systems~\cite{Braginsky_Quantum_measurement,Grangier_Nature_1998}.
It couples two quantum systems so that a signal observable of one system is measured without disturbance by measuring a probe observable of the other system.
This is also the essence of error syndrome identification in general quantum error correction,
including the special case of continuous-variable (CV) error correction schemes~\cite{Braunstein_PRL_1998,Lloyd_PRL_1998,Gottesman_PRA_2001}.
In the context of CV quantum information processing, a QND interaction is also referred to as a sum gate,
transforming the position quadratures of two modes as
$\hat{x}_1\to\hat{x}_1$ and $\hat{x}_2\to\hat{x}_2+\hat{x}_1$ in the Heisenberg picture~\cite{Braunstein_RMP_2005,vanLook_Furusawa_2011}.
This is a direct analogue of a controlled-NOT gate for qubits
and considered as an elementary two-mode entangling gate for universal quantum computation over continuous variables~\cite{Bartlet_PRL_2002,Lloyd_PRL_1999}.
The QND gate is an integral part of many important CV quantum information protocols,
such as generation of cluster states for one-way quantum computing~\cite{Zhang_PRA_2006,Menicucci_PRL_2010},
realization of non-Gaussian gates via gate teleportation~\cite{Gottesman_PRA_2001,Petr_arxiv_2017},
as well as CV coherent communication~\cite{Wilde_PRA_2007}.

Thus far, QND interactions and measurements have been demonstrated in several optical experiments.
One scheme to implement the QND gate is to directly couple two input optical fields and pump fields via parametric amplification using a nonlinear optical media~\cite{Pereira_PRL_1994}.
However, this direct scheme typically induces coupling losses on fragile quantum states, degrading the gate fidelity.
Another scheme which does not require direct coupling and instead uses linear optics and offline-prepared squeezed states
was proposed~\cite{Filip_PRA_2005} and demonstrated with high precision~\cite{Yoshikawa_PRL_2008}.
This offline scheme is, in principle, applicable to arbitrary optical quantum states.
However, previous QND gates~\cite{Yoshikawa_PRL_2008, Ukai_PRL_2011, Yokoyama_PRA_2014} only work on quantum states in narrow sideband modes in the frequency domain.
That is, they are not applicable to general non-Gaussian quantum states generated in wave-packet modes,
such as single-photon states~\cite{Lvovsky_PRL_2001,Neergaard_OptExp_2007} and Schr{\"o}dinger's cat states~\cite{Grangier_Science_2006,Neergaard_PRL_2006},
although such states are included in many CV protocols and are also required for universal quantum computing~\cite{Gottesman_PRA_2001,Bartlet_PRL_2002,Lloyd_PRL_1999}.

In this Letter, we demonstrate QND gate operations and measurements in real time on continuously fluctuating signals with a bandwidth of about 100~MHz.
Unlike previous experiments~\cite{Yoshikawa_PRL_2008, Ukai_PRL_2011, Yokoyama_PRA_2014}, the input signal is randomly fluctuating with a short autocorrelation time, and thus the success of QND interactions on this signal is a proof that our gate correctly operates instant signals without memory-like effects. 
The time-domain traces of quadrature values are obtained in real time by just applying electric filters~\cite{Ogawa_PRL_2016}, and thus can be interpreted as results of real-time QND measurements with respects to time-shifted wave-packet modes determined by the electric filters. 
Since our QND gate works on any wave-packet modes for up to about 100~MHz, our gate is compatible with non-Gaussian quantum states localized in a wave-packet mode. 
Note that, for CV single-mode squeezing and teleportation gates, the bandwidth has been widened to about 10~MHz and operations on non-Gaussian quantum states have already been  demonstrated~\cite{Lee_Science_2011,Takeda_Nature_2013,Miwa_PRL_2014,Miyata_PRA_2014,Fuwa_PRL_2014,Takeda_PRL_2015}.
Here we demonstrate for the first time a broadband {\it interaction} gate, and furthermore the bandwidth is widened to about 100~MHz. 
Our gate is a crucial component for future realizations of non-Gaussian gates~\cite{Gottesman_PRA_2001,Petr_arxiv_2017}, time-domain multiplexed cluster states~\cite{Menicucci_PRL_2010},
error syndrome measurements of a qubit encoded in an oscillator~\cite{Gottesman_PRA_2001}, and CV gate sequences in a loop-based architecture~\cite{Takeda_PRL_2017}.   

{\it Theory}.---Let us define quadratures of a quantum optical field mode $k$ as $\hat{x}_k$ and $\hat{p}_k$ with
$\hat{x}_k\equiv(\hat{a}^{\dag}_k+\hat{a}_k)$ and $\hat{p}_k\equiv i(\hat{a}^{\dag}_k-\hat{a}_k)$, where $\hat{a}_k$ and $\hat{a}^{\dag}_k$ are annihilation and creation operators, respectively
($\hbar=2$, $[\hat{x}_k,\ \hat{p}_{k'}]=2i\delta_{kk'})$.
The QND interaction is a two-mode unitary operation
$\hat{U}_{\rm QND}=\exp\left(-\frac{i}{2}G\hat{x}_1\hat{p}_2\right)$,
where $G$ is a QND gain, i.e. the strength of the interaction of two optical modes.
This interaction transforms the quadrature operators as
\begin{align}
\begin{bmatrix}
\hat{x}_1^{\rm out}\\
\hat{x}_2^{\rm out}
\end{bmatrix}
&=
\begin{bmatrix}
1&0\\
G&1
\end{bmatrix}
\begin{bmatrix}
\hat{x}_1^{\rm in}\\
\hat{x}_2^{\rm in}
\end{bmatrix},&
\begin{bmatrix}
\hat{p}_1^{\rm out}\\
\hat{p}_2^{\rm out}
\end{bmatrix}
&=
\begin{bmatrix}
1&-G\\
0&1
\end{bmatrix}
\begin{bmatrix}
\hat{p}_1^{\rm in}\\
\hat{p}_2^{\rm in}
\end{bmatrix}.
{\label{eq:ideal_input_output}}
\end{align}

\begin{figure}
\includegraphics[width=8.5cm]{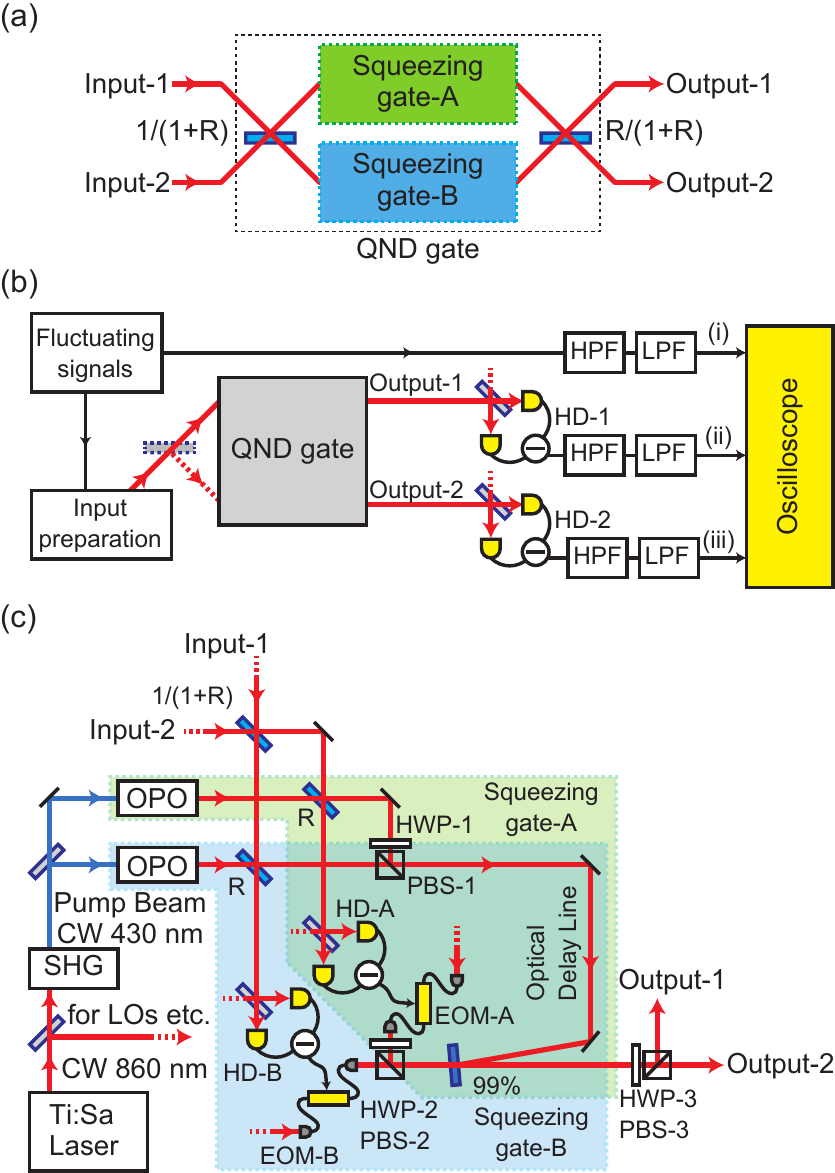}
\caption{
\label{fig:setup}
Experimental setup. 
(a) Decomposition of a QND gate. 
(b) Peripheral systems for the input signal preparation and the output measurements.
Input optical signal is sent to either Input-1 or Input-2. 
(c) Experimental setup of the QND gate. 
Squeezing gate-A and B share an optical delay line in orthogonal polarizations.
SHG, second harmonic generator; HD, homodyne detector.}
\end{figure}

Since the QND interaction belongs to the class of Gaussian operations, it is decomposable into
beam-splitter interactions and single-mode squeezing operations [See Fig.~\ref{fig:setup}(a)]~\cite{Braunstein_PRA_2005, Filip_PRA_2005}.
Furthermore, squeezing is also a Gaussian operation, and is realized by an offline scheme with a beam splitter and
ancillary squeezed light~\cite{Filip_PRA_2005}, where the squeezing degree is tunable via the reflectivity of the beam splitter $R$.
The QND gate is implemented by choosing the beam-splitter reflectivities before and after the squeezing gates as $1/(1+R)$ and $R/(1+R)$, respectively~[See Fig.~\ref{fig:setup}(a)].
We obtain
\begin{subequations}
{\label{eq:experimental_input_output}}
\begin{align}
\hat{x}_1^{\rm out}&=\hat{x}_1^{\rm in}-\sqrt{\frac{1-R}{1+R}}\ \hat{x}_{\rm A}^{(0)}e^{-r_{\rm A}},\\
\hat{x}_2^{\rm out}&=\frac{1-R}{\sqrt{R}}\hat{x}_1^{\rm in}+\hat{x}_2^{\rm in}+\sqrt{R\frac{1-R}{1+R}}\ \hat{x}_{\rm A}^{(0)}e^{-r_{\rm A}},\\
\hat{p}_1^{\rm out}&=\hat{p}_1^{\rm in}-\frac{1-R}{\sqrt{R}}\hat{p}_2^{\rm in}+\sqrt{R\frac{1-R}{1+R}}\ \hat{p}_{\rm B}^{(0)}e^{-r_{\rm B}},\\
\hat{p}_2^{\rm out}&=\hat{p}_2^{\rm in}+\sqrt{\frac{1-R}{1+R}}\ \hat{p}_{\rm B}^{(0)}e^{-r_{\rm B}},
\end{align}
\end{subequations}
where $\hat{x}_{\rm A}^{(0)}e^{-r_{\rm A}}$ and $\hat{p}_{\rm B}^{(0)}e^{-r_{\rm B}}$ are quadratures of ancillary squeezed vacua of squeezing gates A and B with finite squeezing parameters $r_{\rm A}$ and $r_{\rm B}$. In the ideal limit of $r_{\rm A}, r_{\rm B}\rightarrow\infty$, both $\hat{x}_{\rm A}^{(0)}e^{-r_{\rm A}}$ and $\hat{p}_{\rm B}^{(0)}e^{-r_{\rm B}}$ terms vanish, and Eq.~\eqref{eq:experimental_input_output} becomes equivalent to Eq.~\eqref{eq:ideal_input_output}, where the QND gain is $G=(1-R)/\sqrt{R}$.
In the experiment, we choose the QND gain $G=1$. In this case, $R=(3-\sqrt{5})/2\approx0.38$, $1/(1+R)\approx0.72$ and $R/(1+R)\approx0.28$.

For characterization of the gate, we obtain the quadratures
$\hat{x}_k$ or $\hat{p}_k$ $(k=1,2)$
by homodyne detection using a local oscillator (LO). 
Generally, in the case that the LO is a continuous coherent light, the detected homodyne signal is also continuous. 
The quadrature of a quantum state in a wave-packet mode $g_{\rm mode}(t)$ is obtained from the original homodyne signal $\hat{X}_{k}(t)$ by an integration  $\hat{x}_{k}=\int g_{\rm mode}(\tau)\hat{X}_{k}(\tau)d\tau$. 
On the other hand, when a continuous signal $\hat{X}_{k}(t)$ passes through a filter with a response function $g_{\rm filter}(t)$, the resulting continuous signal becomes $\hat{x}_{k}(t)=\int g_{\rm filter}(t-\tau)\hat{X}_{k}(\tau)d\tau$. 
Therefore we obtain quadrature values in real time just by inserting an electric filter, where the mode function that corresponds to the quadrature value $\hat{x}_{k}(t_0)$ obtained at time $t_0$ is $g_{{\rm mode},t_0}(t)=g_{\rm filter}(t_0-t)$~\cite{Ogawa_PRL_2016}.
Note that real-time measurements are necessary for nonlinear feed-forward operations in measurement-based quantum computation~\cite{Ogawa_PRL_2016}.   
We choose a low-pass filter (LPF) which has a flat passband and a steep edge with a cutoff frequency of 100 MHz in order to treat the bandwidth of 100 MHz equally. 
However, the QND gate itself can work on arbitrary wave-packet modes for up to the bandwidth of 100~MHz, enabling operations on non-Gaussian states.

As already noted, in order to show memoryless features of our gate, we use random white signals as inputs. 
From the signal-to-noise ratio of this random signal, we can evaluate the conventional QND quantities $T_{\rm S}$ and $T_{\rm P}$~\cite{Holland_PRA_1990}.
However, unlike previous experiments~\cite{Grangier_Nature_1998,Pereira_PRL_1994,Yoshikawa_PRL_2008, Ukai_PRL_2011, Yokoyama_PRA_2014}, it may not be appropriate to evaluate $T_{\rm S}$ and $T_{\rm P}$ just by transfer of signal powers.
If the signal is modified unexpectedly by irregular gate responses, a part of the input signal is considered to be converted to noise at the output, by which the effective $T_{\rm S}$ and $T_{\rm P}$ degrades.
In order to exclude such a possibility, we check the cancellation of the output signals by using the input signal.
The setup is shown in Fig.~\ref{fig:setup}(b).
The random signal is split into two; one is utilized for generating the input optical signal, and the other is stored for reference.
Here we set the target of the QND measurement to the quadrature amplitude produced by the random signal in the wave-packet mode defined by the electric filters.
This input amplitude is directly stored by applying the same electric filters to the random signal before storage.
Therefore, we can cancel the produced output signals [(ii) and (iii) in Fig.~\ref{fig:setup}(b)] by using the stored signal [(i) in Fig.~\ref{fig:setup}(b)] with an appropriate shift of the time origin.
This is also a new achievement of this research. 

{\it Experimental setup}.---We use a continuous-wave (CW) Ti:Sa laser at a wavelength of 860 nm. 
Input states of the QND gate are vacuum states and coherent states.
We generate a random optical signal using a waveguide electro-optics modulator (EOM) and an amplified Johnson electric noise, which is applied 
to each of the input quadratures $(x_1^{\rm in},x_2^{\rm in},p_1^{\rm in},p_2^{\rm in})$.
For the frequency characteristic of the random signal and the scheme of generating the coherent state, see the Supplemental Material (SM)~\cite{Supplementary}.
The other three input quadratures are at vacuum levels.
This is sufficient to characterize the gate-response matrix on the assumption of the linearity of the gate. 

The QND gate consists of a Mach-Zehnder interferometer containing two squeezing gates in it as shown in Fig.~\ref{fig:setup}(a). 
The squeezing gate has an optical delay line to compensate the delay of electronic circuits for feed-forward operations. 
In order to match the delays of two squeezing gates, we implement a common delay line (about 3~m) by  utilizing the optical polarization degrees of freedom as shown in Fig.~\ref{fig:setup}(c). 
We insert a half-wave plate (HWP) before a polarizing beam splitter (PBS) to separate the two outputs, by which the latter beam splitter $R/(1+R)$ is implemented. 
The ancillary squeezed vacua are generated from triangle-shaped optical parametric oscillators (OPOs)~\cite{Serikawa_Opt_Exp_2016}. 
For the broadband spectra of ancillary squeezed vacua and homodyne detectors, see the SM~\cite{Supplementary}. 

We apply, in addition to the 100-MHz LPF mentioned above, a high-pass filter (HPF) with a cutoff frequency of 1~MHz to the output homodyne signals for rejection of low-frequency noises. 
The mode function is mainly determined by the LPF, and the deformation of it by the HPF is negligible. 
The frequency characteristic of these filters are shown in the SM~\cite{Supplementary}.
We acquire the filtered homodyne signals, together with the filtered input signal, by an oscilloscope at the sampling rate of 1 GHz. 
For the QND quantities $T_{\rm S}$, $T_{\rm P}$ and $V_{\rm S|P}$~\cite{Holland_PRA_1990}, we use 1,000 sets of sequential 10,000 data points.  
For the power spectra, we use 9,000 sets of sequential 1,024 data points.

\begin{figure}
\includegraphics[width=8.5cm]{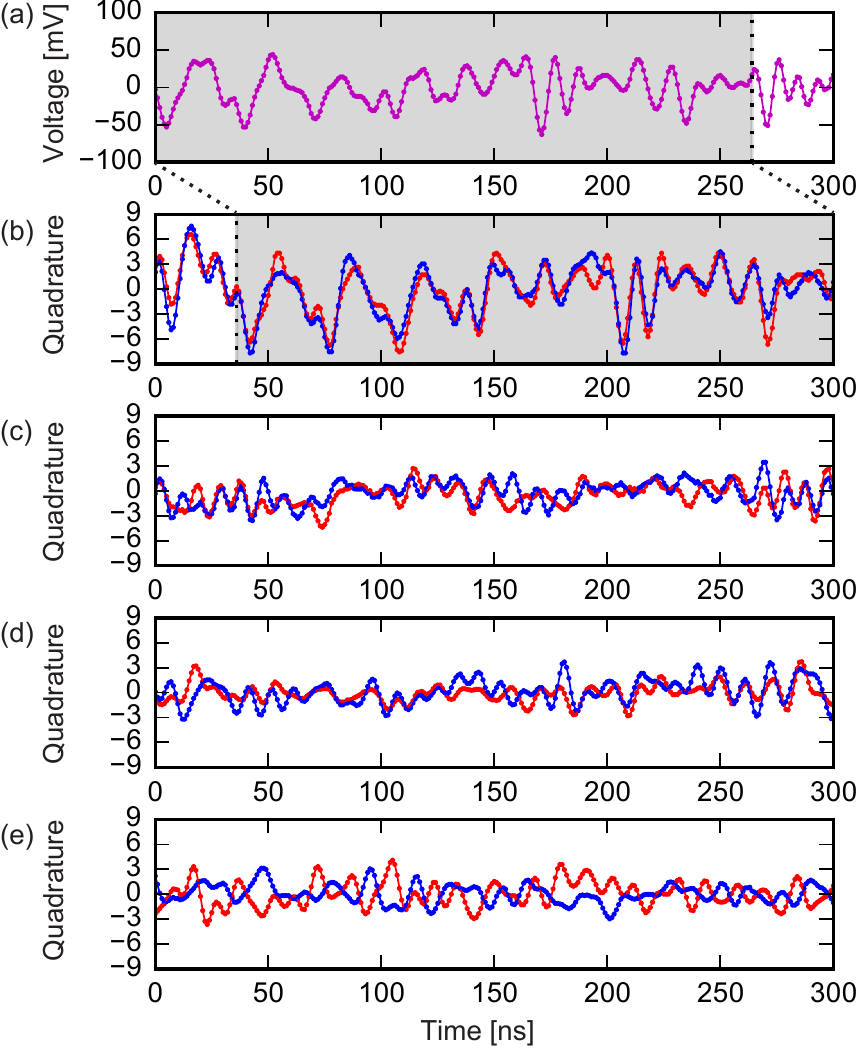}
\caption{\label{fig:realtime_result}Time-domain traces for 300~ns.
(a) The filtered input white signals. 
(b) The filtered output homodyne signals of $\hat{x}_1^{\rm out}$ (red) and $\hat{x}_2^{\rm out}$ (blue). 
(c) Results of the cancellation.
(d) The filtered output homodyne signals of $\hat{x}_1^{\rm out}$ (red) and $\hat{x}_2^{\rm out}$ (blue) for vacuum inputs. 
(e) The filtered output homodyne signals of $\hat{p}_1^{\rm out}$ (red) and $\hat{p}_2^{\rm out}$ (blue) for vacuum inputs. 
}
\end{figure}

\begin{figure*}
\includegraphics[width=17cm]{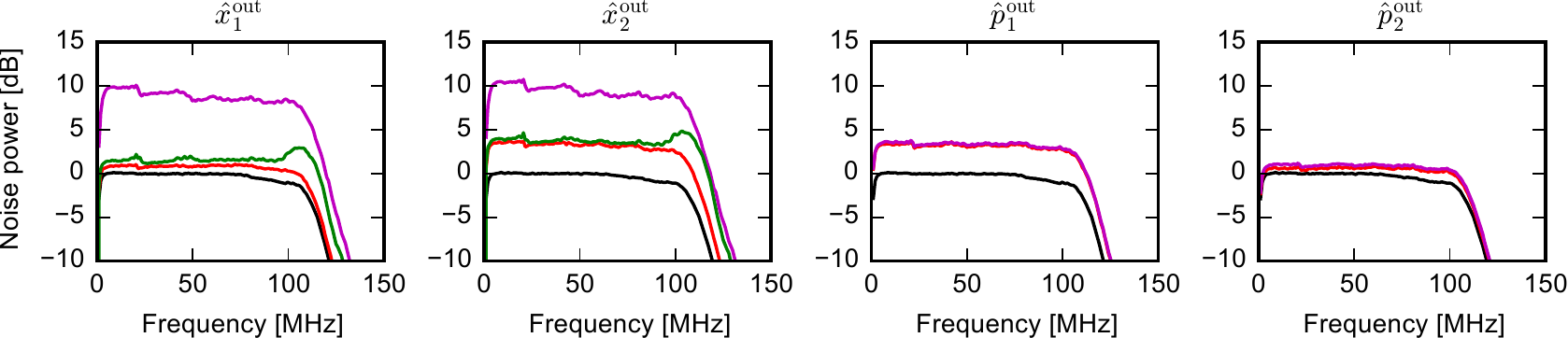}
\caption{
\label{fig:cancel}
Power spectra of output quadratures when a random signal is added to $\hat{x}_1^{\rm in}$. 
Black: shot noises. 
Red: results of vacuum-state input. 
Magenta: results of coherent-state input. 
Green: cancellation of the random signal.}
\end{figure*}

{\it Experimental results}.---First, as an example, we show the time-domain traces for the case where the white signal is applied to $\hat{x}_1^{\rm in}$.
The other three cases are shown in the SM~\cite{Supplementary}.
In Fig.~\ref{fig:realtime_result}, we show typical time-domain traces of the filtered white signals and the filtered homodyne signals for 300~ns. 
Figures~\ref{fig:realtime_result}(a) and (b) show the traces of the input white signal and the output quadratures $\hat{x}_1^{\rm out}$ and $\hat{x}_2^{\rm out}$ [(i), (ii), and (iii) in Fig.~\ref{fig:setup}(b)], respectively.
We can see that the output quadratures $\hat{x}_1^{\rm out}$ and $\hat{x}_2^{\rm out}$ follow the input white signal with a time delay of 36~ns, which is shown by gray backgrounds and dotted lines in Figs.~\ref{fig:realtime_result}(a) and (b).
This means that the signal input $\hat{x}_1^{\rm in}$ is transmitted non-destructively to the signal output $\hat{x}_1^{\rm out}$, and simultaneously the signal information is copied to the probe output $\hat{x}_2^{\rm out}$.
Then we subtract the input white signal from the output respective quadratures $\hat{x}_1^{\rm out}$ and $\hat{x}_2^{\rm out}$ with an optimum gain and the time shift; the results are shown in Fig.~\ref{fig:realtime_result}(c). 
As references, in Fig.~\ref{fig:realtime_result}(d), we also show traces of $\hat{x}_1^{\rm out}$ and $\hat{x}_2^{\rm out}$ for the case of vacuum input.
We can see that the variances of the residual fluctuations in Fig.~\ref{fig:realtime_result}(c) are comparable to those of the vacuum input case in Fig.~\ref{fig:realtime_result}(d).
The nice cancellation with a simple time shift means that the gate converts the instant input signals to the instant output signals without memory-like effects in this time scale.
Without the added random signals, there is still some positive correlation independent of the input signal in $\hat{x}_1^{\rm out}$ and $\hat{x}_2^{\rm out}$.
On the other hand, when we look at $\hat{p}_1^{\rm out}$ and $\hat{p}_2^{\rm out}$ in Fig.~\ref{fig:realtime_result}(e), there is a negative correlation.
Figures~\ref{fig:realtime_result}(d) and (e) show the quantum entanglement generated by the gate interaction.

Next, in order to evaluate the cancellation more precisely, we perform Fourier transform to the results, and the resulting power spectra are shown in Fig.~\ref{fig:cancel}. 
The spectra for the vacuum-state input, the coherent-state input, the cancellation, and the homodyne shot noise as a reference are colored in red, magenta, green, and black, respectively. 
In the case of an ideal QND interaction of vacuum inputs with $G=1$, $\hat{x}_1^{\rm out}$ and $\hat{p}_2^{\rm out}$ are kept at the shot-noise level, while $\hat{x}_2^{\rm out}$ and $\hat{p}_1^{\rm out}$ are increased by 3~dB from the shot-noise level, because a vacuum fluctuation of $\hat{x}_1^{\rm in}$ or $\hat{p}_2^{\rm in}$ is added.
Our results are in good agreement with this, though there are some excess noise increases due to finite squeezing of ancillary states.
When the input white signal is added to $\hat{x}_1^{\rm in}$, the powers of $\hat{x}_1^{\text{out}}$ and $\hat{x}_2^{\text{out}}$ increase by the same amount, showing the unity gain of the QND interaction, while those of $\hat{p}_1^{\text{out}}$ and $\hat{p}_2^{\text{out}}$ do not increase, showing negligible crosstalk between $x$ and $p$ quadratures.
Comparing the vacuum-input (red) trace and the signal-canceled (green) trace, we can see that the cancellation is almost perfectly working for up to about 100~MHz.
Further discussions of the cancellations by introducing response functions are included in the SM~\cite{Supplementary}.

\begin{figure}
\includegraphics[width=8.5cm]{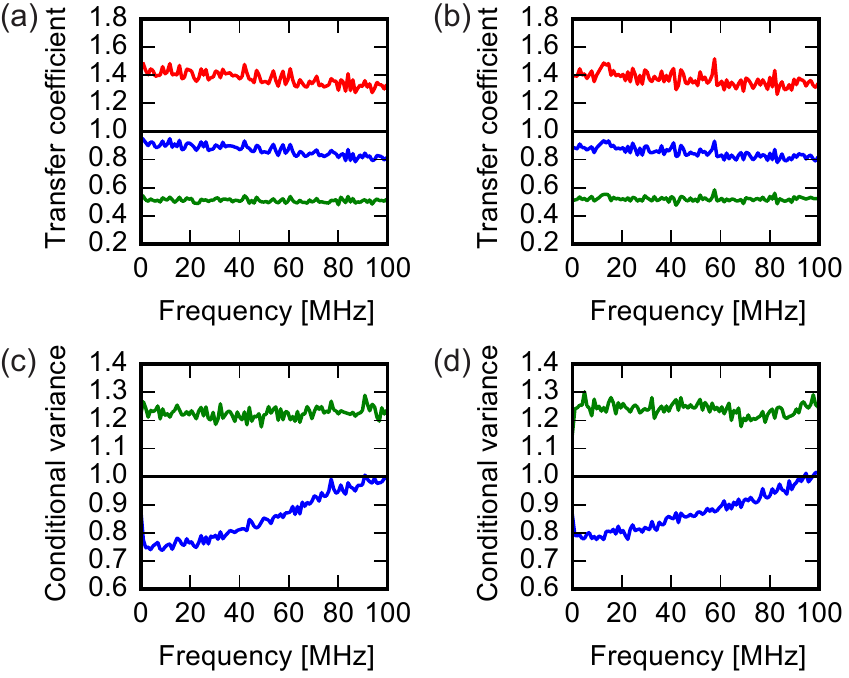}
\caption{
\label{fig:Transfer_coefficients_cond_var}
Spectra of the QND quantities.
(a), (b) Transfer coefficients when the random signal is added to $\hat{x}_1^{\rm in}$ or $\hat{p}_2^{\rm in}$, respectively. 
Blue, green and red traces are $T_{\rm S}$, $T_{\rm P}$ and $T_{\rm S}+T_{\rm P}$, respectively. 
(c), (d) Conditional variances for vacuum inputs, i.e., the variances of $\hat{x}_1^{\rm out}-g_x\hat{x}_2^{\rm out}$ and $g_p\hat{p}_1^{\rm out}+\hat{p}_2^{\rm out}$ at the gain $g_x=0.41$ and $g_p=0.39$, respectively~\cite{Supplementary}, normalized by the shot noise spectrum.
Blue and green traces are for the cases with and without the ancillary squeezed vacua, respectively.
}
\end{figure}

Finally, we evaluate the QND quantities $T_{\rm S}$, $T_{\rm P}$, and $V_{\rm S|P}$ for both $\hat{x}$ and $\hat{p}$ quadratures.
The success of QND measurements is commonly verified by the criteria~\cite{Holland_PRA_1990}
\begin{align}
1&<T_{\rm S}+T_{\rm P},& V_{\rm S|P}&<1.
\end{align}
The experimentally determined values are $T_{\rm S}+T_{\rm P}=1.37\pm0.10>1$ and $V_{\rm S|P}=0.88\pm0.03<1$ for the $\hat{x}$ quadratures, $T_{\rm S}+T_{\rm P}=1.37\pm0.10>1$ and $V_{\rm S|P}=0.88\pm0.03<1$ for the $\hat{p}$ quadratures.
Therefore, we succeeded in construction of a QND gate that enables real-time QND measurements for both conjugate quadratures with the bandwidth of about 100~MHz. 
For a more detailed analysis, we show the QND quantities at each frequency in Fig.~\ref{fig:Transfer_coefficients_cond_var}.
All of $T_{\rm S}$, $T_{\rm P}$, and $V_{\rm S|P}$ satisfy the QND criteria up to about 100 MHz.
As for $V_{\rm S|P}$, because of the finite bandwidth of the ancillary squeezed vacua, the correlation degrades at higher frequencies, however, there are still sub-shot-noise correlations for up to about 100~MHz.
The two output modes are entangled, which is described in the SM~\cite{Supplementary}.

{\it Conclusions}.---We experimentally demonstrated an optical two-mode QND interaction gate that enables real-time QND measurements on temporally fluctuating random signals.
We also showed that the interaction works on a broad spectrum, namely up to about 100 MHz in the frequency domain.
The capability of the gate to deal with instantaneous signals is confirmed by the cancellation of random signals.
This scheme is applicable to any quantum states in wave-packet modes, including non-Gaussian states, and thus
perfectly suitable for implementing non-Gaussian gates~\cite{Gottesman_PRA_2001,Petr_arxiv_2017}, generating 
time-multiplexed cluster states~\cite{Menicucci_PRL_2010}, diagnosing the error syndrome in quantum error correction~\cite{Braunstein_PRL_1998,Lloyd_PRL_1998,Gottesman_PRA_2001}, and implementing gate sequences in a loop-based architecture~\cite{Takeda_PRL_2017}.

\begin{acknowledgments}
This work was partly supported by CREST (JPMJCR15N5) and PRESTO (JPMJPR1764) of JST, JSPS KAKENHI, APSA.
P. v. L. was supported by Qcom (BMBF).
Y. S., S. Y., and T. S. acknowledge financial support from ALPS.
\end{acknowledgments}

\clearpage
\onecolumngrid
\setlength{\parskip}{3mm}

\makeatletter
\renewcommand{\theequation}{S\arabic{equation}}
\renewcommand{\thefigure}{S\arabic{figure}}
\renewcommand{\thetable}{S\arabic{table}}
\renewcommand{\thesection}{S\Roman{section}}
\makeatother





\noindent
\begin{center}
\textbf{\large Supplemental Material for ``Quantum Nondemolition Gate Operations and Measurements in Real Time on Fluctuating Signals''}
\end{center}

\section{\label{sec:Coh_setup}Detailed experimental setup and frequency spectra}

\subsection{Electric filters and response function}
\begin{figure}[b]
\centering
\includegraphics[width=17cm,clip]{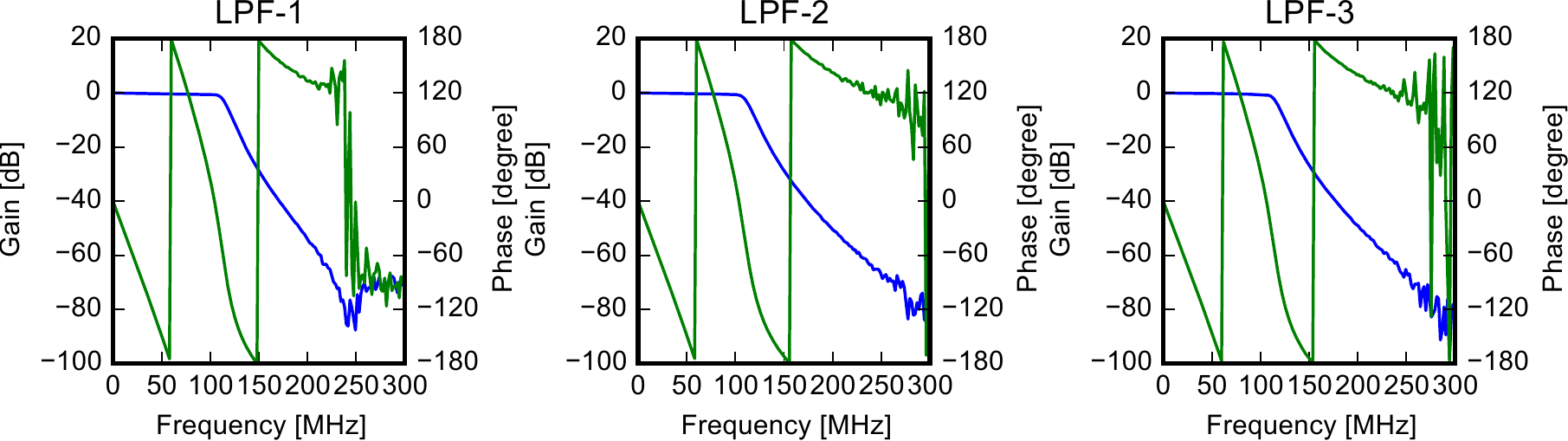}
\caption{\label{fig:LPFs}Frequency characteristics of the LPFs, obtained by a network analyzer (Keysight, E5061B). Blue: gain. Green: phase.}
\end{figure}

\begin{figure}[tb]
\begin{center}
\includegraphics[width=17cm]{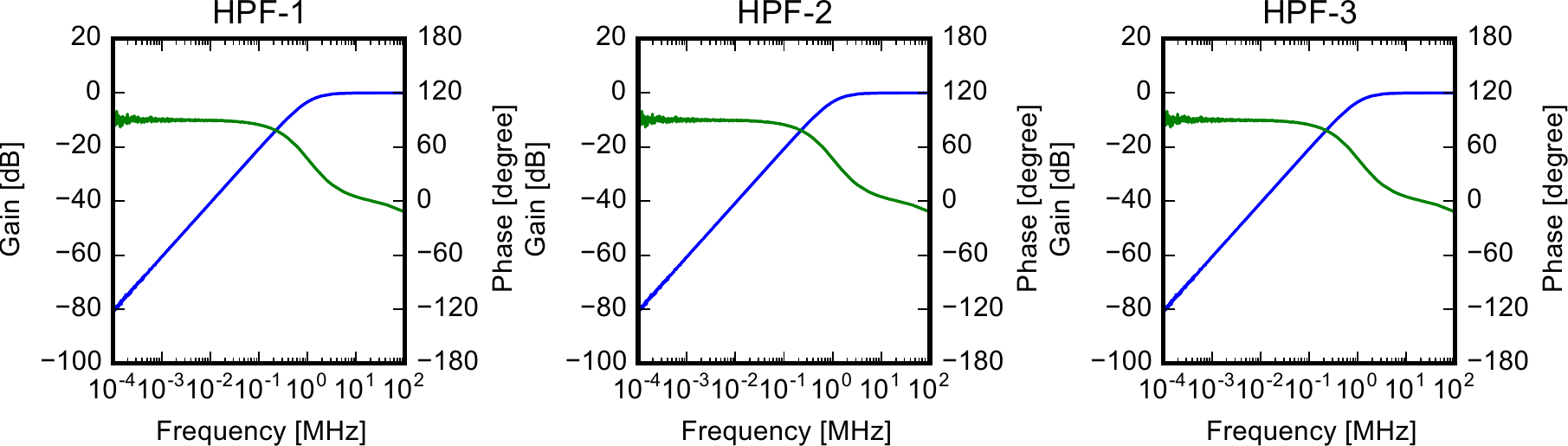}
\caption{\label{fig:HPFs}Frequency characteristics of the HPFs, obtained by a network analyzer (Keysight, E5061B). Blue: gain. Green: phase.}
\end{center}
\end{figure}

\begin{figure}[tb]
\begin{center}
\includegraphics[width=8.5cm]{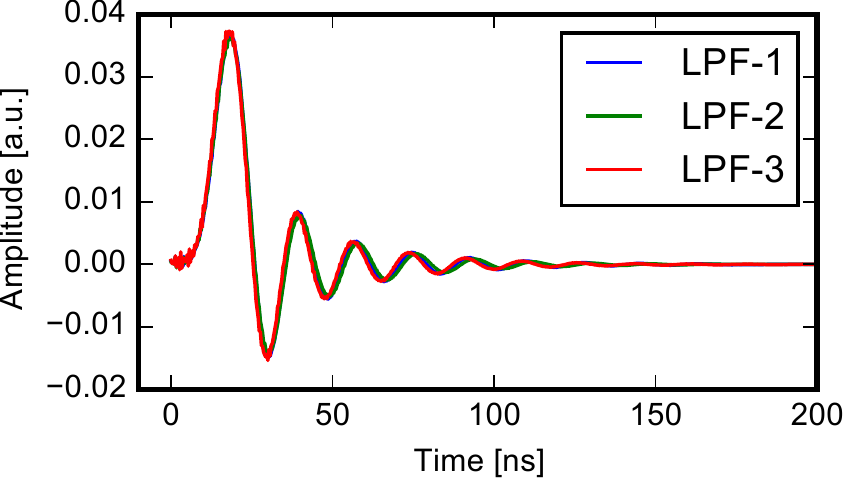}
\caption{\label{fig:mode_function}Obtained response function of the LPF.}
\end{center}
\end{figure}

The homdyne signals for verification as well as the white signals for the QND input are stored by an oscilloscope (DPO7054, Tektronix) after a low-pass filter (LPF) and a high-pass filter (HPF).
The LPF is a commercially available filter whose cut-off frequency is 100 MHz (Mini-Circuits, BLP-100+). 
We plot the frequency characteristics of the LPF in Fig.~\ref{fig:LPFs}. 
The HPF is a homemade 1st-order filter with a cutoff frequency of 1 MHz, which is used in order to remove low-frequency noise around the laser carrier frequency.
We plot the frequency characteristics of the HPF in Fig.~\ref{fig:HPFs}.
The mode function $g_{\text{mode}}(t)$ is mainly determined by the LPF.
The time-domain response function $g_{\rm filter}(t)$ calculated from the gain and phase in Fig.~\ref{fig:LPFs} is shown in Fig.~\ref{fig:mode_function}.
As mentioned in the main text, the time-reversal response function with time shifts is the effective mode function $g_{\rm mode}(t)$ for the QND measurements.

\subsection{White signal source}
\begin{figure}[tb]
\begin{center}
\includegraphics[width=8cm]{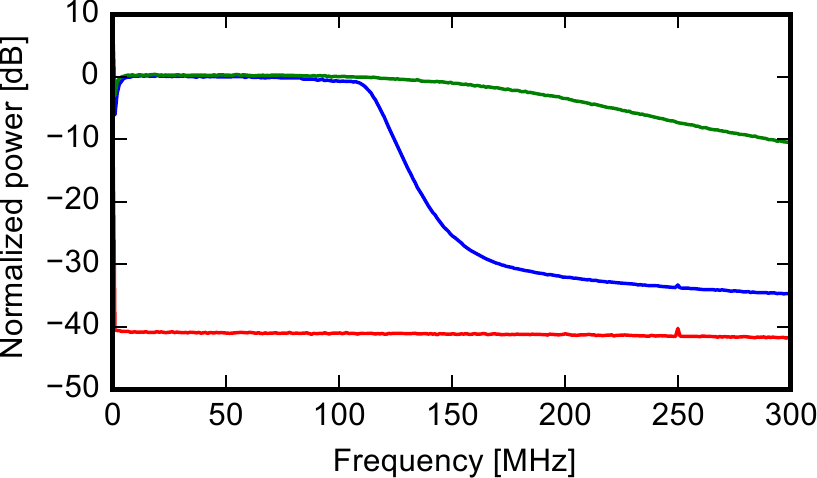}
\caption{\label{fig:whitenoise}
Power spectrum of the input white signal. 
Green: unfiltered signal. 
Blue: filtered signal. 
Red: oscilloscope noise floor.}
\end{center}
\end{figure}

Figure~\ref{fig:whitenoise} shows the power spectrum of the white signal used for the input of the QND gate. 
The white signal is amplified thermal noises of resistors and operational amplifiers (OPA847, Texas Instruments). 
The trace in green represents unfiltered signals, while the trace in blue represents filtered signals which corresponds to the signals stored by the oscilloscope in the actual QND experiment.

\subsection{Homodyne detectors}
\begin{figure}[tb]
\begin{center}
\includegraphics[width=17cm]{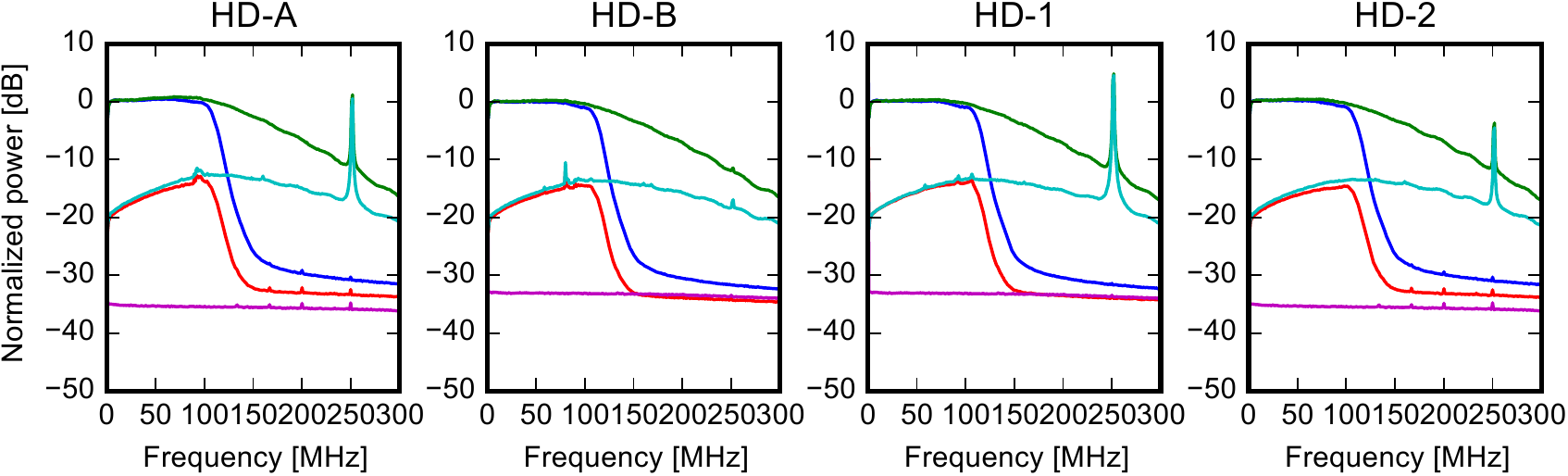}
\caption{\label{fig:SNL}Noise power spectra of four homodyne detectors. 
Blue: optical shot noise spectra with the LPF. 
Green: optical shot noise spectra without the LPF. 
Red: detector dark noise spectra with the LPF. 
Cyan: detector dark noise spectra without the LPF.
Magenta: oscilloscope noise floor.}
\end{center}
\end{figure}

Figure~\ref{fig:SNL} shows the optical shot-noise spectra with a local oscillator (LO) power of 10~mW, together with the detector dark noise spectra, of the four homodyne detectors (two for feed-forward operations and two for QND measurements).
We show both the filtered and unfiltered cases. 
The shot-noise spectra are flat up to about 100~MHz for all of the four detectors. 
The clearance between the shot noise and the dark noise is more than 10~dB even at 100~MHz.

\subsection{Ancillary squeezed vacua}
\begin{figure}[tb]
\begin{center}
\includegraphics[width=11cm]{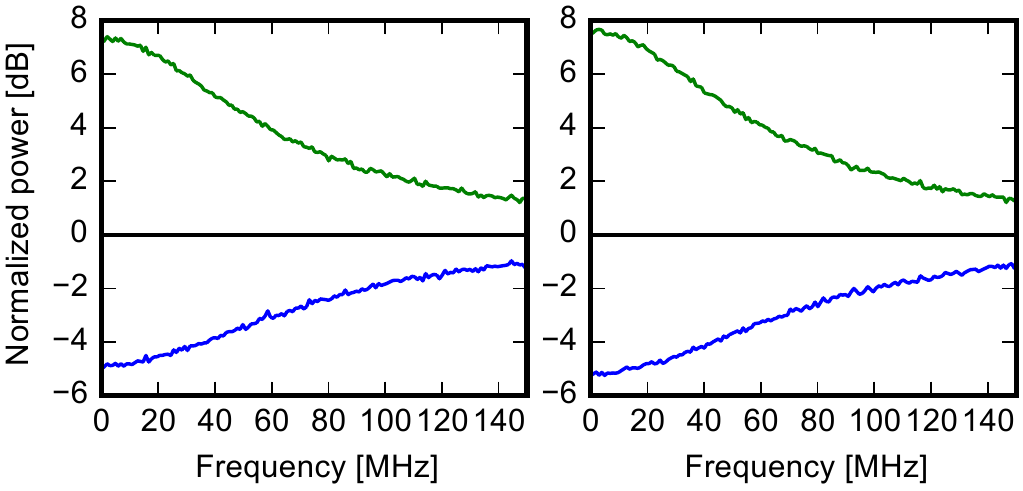}
\caption{\label{fig:squeeze}
Noise power spectra of ancillary squeezed vacua, normalized by the shot-noise spectrum.
Green: anti-squeezed quadrature.
Blue: squeezed quadrature. 
}
\end{center}
\end{figure}

Figure~\ref{fig:squeeze} shows the power spectra of the squeezed and anti-squeezed quadratures of the ancillary squeezed vacua normalized by the shot noise spectrum.
The power of the pump beam is 85~mW.
Both of the two squeezed vacua show about $-5$~dB of squeezing at low frequencies and about $-2$~dB of squeezing at 100 MHz.
These spectra are in good agreement with the bandwidths of the OPO cavities (about 150~MHz of full width at half maximum).

\begin{figure}[tb]
\centering
\includegraphics[width=10cm]{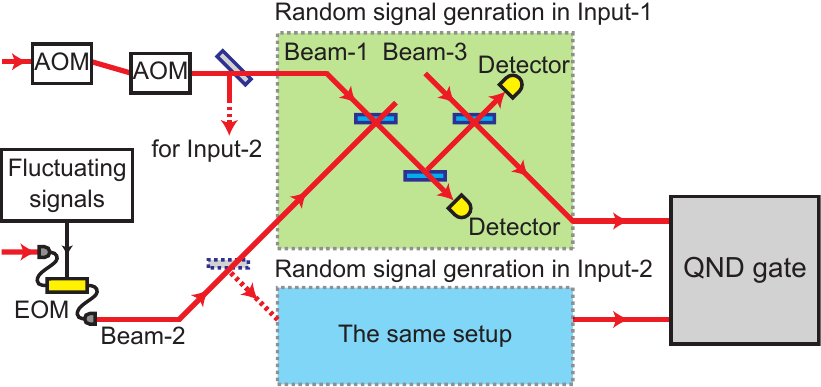}
\caption{\label{fig:Coh_setup}
Experimental setup for input signal preparation. Beam-2 is sent to either Input-1 or Input-2.}
\end{figure}

\subsection{Control of optical systems}
In order to lock interference phases in the QND gate, we use weak laser beams as phase references for each optical paths.
These reference beams are temporally turned on and off by switching a pair of acousto-optic modulators (AOMs).
We control the optical systems by feedback when the reference beams are on, while the system is held and the QND gate is tested when they are off.
The duration of ON time is 1400~$\mu$s and that of OFF time is 600~$\mu$s.
The QND measurement data are acquired within 10~$\mu$s in the OFF time, during which the drift of the optical system is negligible.

However, there are some beams which cannot be turned off.
Some are carrier beams to generate input random signals by modulations, and others are carrier beams for feed-forward operations in the squeezing gates.
The laser noises of these beams disturb the homodyne signals.
The noises by the input carrier beams are more significant than those by the feed-forward carrier beams because of the differences in optical path lengths.
Since the input beams pass through the optical delay line before interference with the LOs, the phase noises look larger in the output homodyne signals.
These noises are filtered out by the HPF in Fig.~\ref{fig:HPFs} and thus they are not so significant problems, however, in order to further remove them, we employ procedures as follows.

The optical setup for the input random signal is shown in Fig.~\ref{fig:Coh_setup}. 
For each of Input-1 and Input-2, three beams are used. 
Beam-1 is the phase reference for the QND gate. 
Beam-2 is the carrier beam to convert the random electronic signals to optical signals by phase modulation. 
While Beam-1 is temporally turned off during data acquisition, Beam-2 is  always on.
We use Beam-3, which is always on, for canceling the carrier component of Beam-2, leaving only the modulation sideband.  
Note that only Beam-1 is used when we use vacuum states as input states. 
The quadrature to add the random signal is selected via the relative phase between Beam-1 and Beam-2. 
For example, for Input-1, when Beam-1 and Beam-2 are locked in phase, the random signal is added to $\hat{x}_1^{\rm in}$. 
On the other hand, when Beam-1 and Beam-2 are locked 90-degree out of phase, the random signal is added to $p_1^{\rm in}$. 
Beam-3 is always locked to the opposite phase with Beam-2, removing the carrier component of Beam-2.

\subsection{Feed-forward operation}
The feed-forward operations in the squeezing gates cancel the anti-squeezed noises of the ancillary squeezed vacua.
Here we explain this by using equations.
In the squeezing gate, first the input state (quadrature operators $\hat{x}^{\rm in}$ and $\hat{p}^{\rm in}$) is coupled with an ancillary squeezed state (quadrature operators $\hat{x}^{(0)}e^{-r}$ and $\hat{p}^{(0)}e^{r}$ with a squeezing parameter $r$) by a beam splitter with a reflectivity $R$.
\begin{subequations}
\begin{align}
\hat{x}^{\text{int-1}}&=\sqrt{R}\hat{x}^{\rm in}+\sqrt{1-R}\hat{x}^{(0)}e^{-r},&
\hat{p}^{\text{int-1}}&=\sqrt{R}\hat{p}^{\rm in}+\sqrt{1-R}\hat{p}^{(0)}e^{r},\\
\hat{x}^{\text{int-2}}&=\sqrt{1-R}\hat{x}^{\rm in}-\sqrt{R}\hat{x}^{(0)}e^{-r},&
\hat{p}^{\text{int-2}}&=\sqrt{1-R}\hat{p}^{\rm in}-\sqrt{R}\hat{p}^{(0)}e^{r}.
\end{align}
\end{subequations}
Next, as a feed-forward operation, the anti-squeezed quadrature of a beam-splitter output $\hat{p}^{\text{int-2}}$ is measured and used for cancellation of the anti-squeezed noise $\hat{p}^{(0)}e^{r}$ in the other output quadrature $\hat{p}^{\text{int-1}}$,
\begin{align}
\hat{x}^{\text{out}}&=\hat{x}^{\text{int-1}}=\sqrt{R}\hat{x}^{\rm in}+\sqrt{1-R}\hat{x}^{(0)}e^{-r},&
\hat{p}^{\text{out}}&=\hat{p}^{\text{int-1}}+\sqrt{\frac{1-R}{R}}\hat{p}^{\text{int-2}}=\frac{1}{\sqrt{R}}\hat{p}^{\rm in}.
\label{eq:squeezing _gate}
\end{align}
In the ideal limit of $r\rightarrow\infty$, the excess noise term $\hat{x}^{(0)}e^{-r}$ vanishes, and Eq.~\eqref{eq:squeezing _gate} approaches the ideal squeezing transformation where the squeezing degree is determined by the reflectivity $R$.

\begin{figure}[tb]
\centering
\includegraphics[width=17cm]{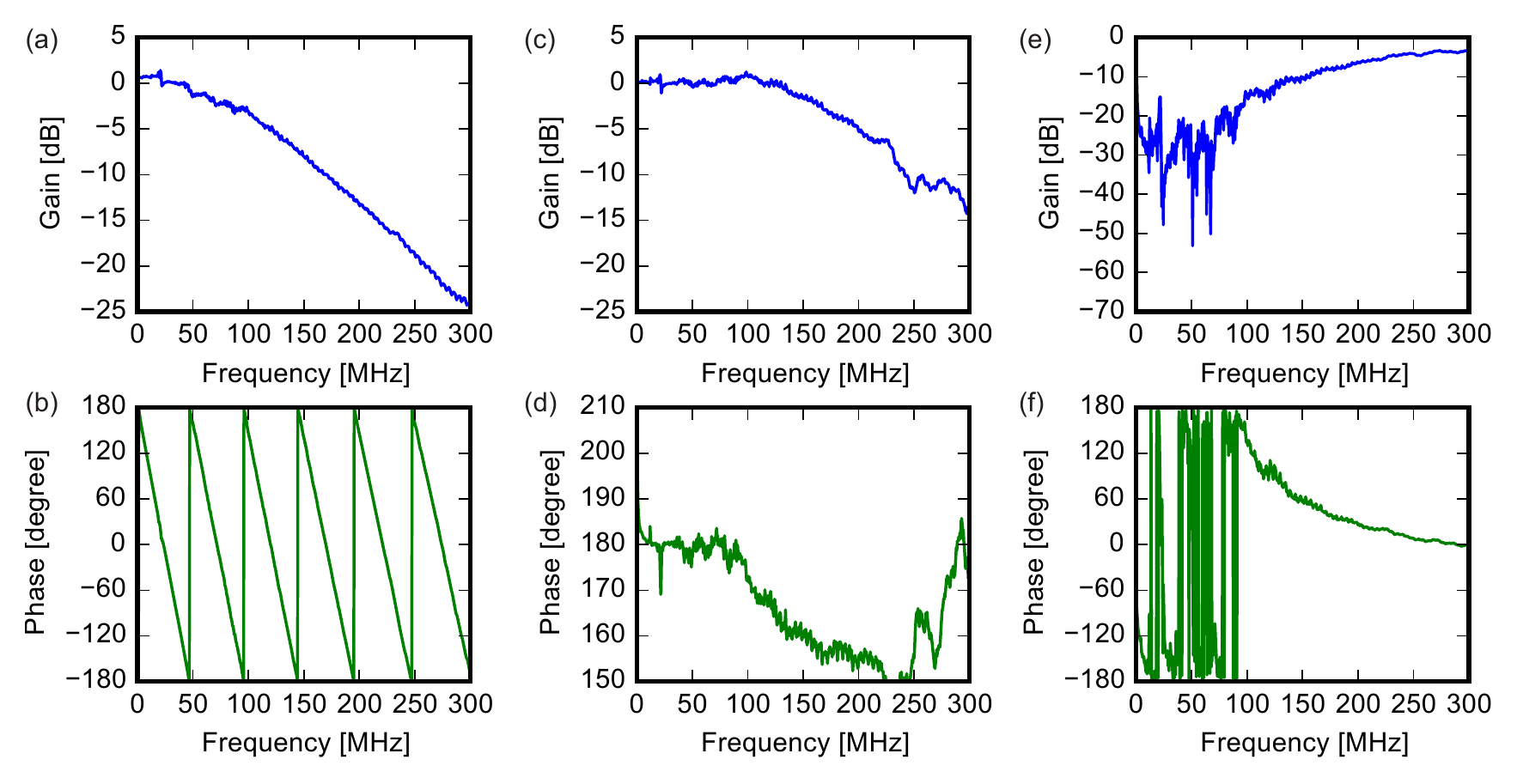}
\caption{\label{fig:CancelA}
Cancellation of the modulated signal by the feed-forward in the squeezing gate-A.
(a), (b) Gain and phase of the reference beam in the squeezing gate-A, used for calibration of the other traces.
(c), (d) Gain and phase of the feed-forward beam in the squeezing gate-A.
(e), (f) The results of the cancellation of the modulated signal.}
\end{figure}

\begin{figure}[tb]
\centering
\includegraphics[width=17cm]{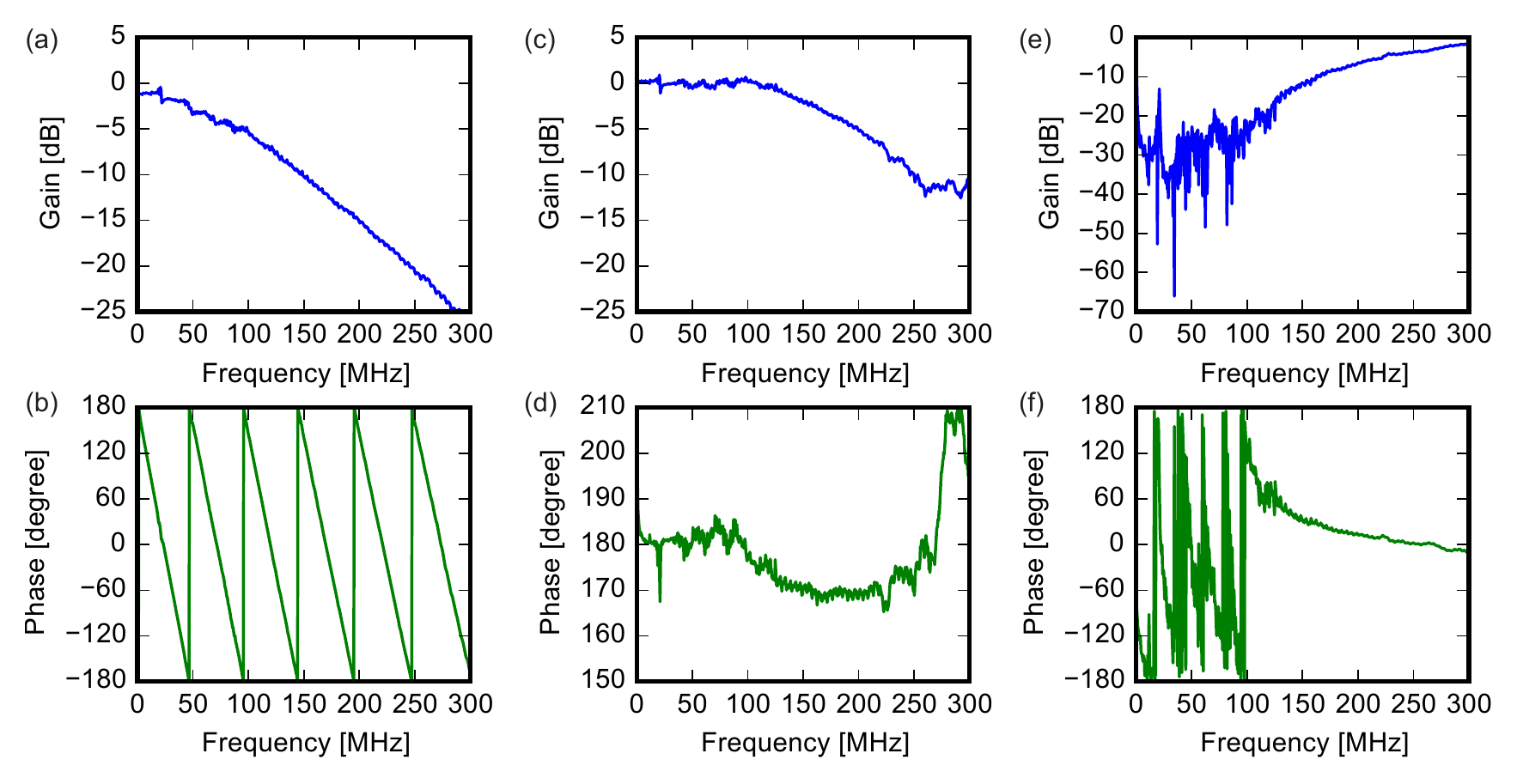}
\caption{\label{fig:CancelB}
Cancellation of the modulated signal by the feed-forward in the squeezing gate-B.
(a), (b) Gain and phase of the reference beam in the squeezing gate-B, used for calibration of the other traces.
(c), (d) Gain and phase of the feed-forward beam in the squeezing gate-B.
(e), (f) The results of the cancellation of the modulated signal.}
\end{figure}

For the cancellation of the anti-squeezed noises, unlike the previous narrowband experiments~\cite{A-Yoshikawa_PRL_2008,A-Ukai_PRL_2011}, the electronic signal for the feed-forward must be synchronized with the optical signal, in other words, the phase lags must be matched at all the frequencies.
For this purpose, we use high-speed homodyne detectors and amplifiers with a flat gain and a linear dispersion, and the optical delay line for the compensation of the electronic delay.
We confirmed the broadband cancellation by using a network analyzer (MS4630B, ANRITSU), which is shown in Figs.~\ref{fig:CancelA} and \ref{fig:CancelB}.
Modulation signals are added by an EOM before the OPOs to the ancillary quadratures to be anti-squeezed, and they are canceled by the feed-forward.
Figures~\ref{fig:CancelA}(a), ~\ref{fig:CancelA}(b), ~\ref{fig:CancelB}(a), and \ref{fig:CancelB}(b) are the gains and phases of the modulated reference beams through the optical delay line.
The gains decrease at higher frequencies due to the bandwidth of the OPO cavities.
They are used for calibration of the traces in the other figures in Figs.~\ref{fig:CancelA} and \ref{fig:CancelB}.
Figures~\ref{fig:CancelA}(c), ~\ref{fig:CancelA}(d), ~\ref{fig:CancelB}(c), and \ref{fig:CancelB}(d) are the gains and phases through the feed-forward electronic paths.
The gains are flat and the phases are opposite (180$^\circ$) for up to 100~MHz.
Figures~\ref{fig:CancelA}(e), ~\ref{fig:CancelA}(f), ~\ref{fig:CancelB}(e), and \ref{fig:CancelB}(f) are the residual modulation signals after the cancellation.
The extinction ratios of the modulated signals are more than 20~dB for up to 100~MHz.

\section{\label{sec:transfer_func}Response of the QND gate}
\subsection{\label{subsec:general_theory}General theory of response functions and cancellation}
We consider a linear and static system
\begin{align}
y(t)=\int f(t-\tau)w(\tau)d\tau+v(t), 
\label{eq:transfer_1}
\end{align}
where $f(t)$ is a response function, $w(t)$ is an input signal, $y(t)$ is an output signal, and $v(t)$ is an excess noise which is independent of $w(t)$, i.e., the cross-correlation vanishes,
\begin{align}
R_{wv}(t)=\Braket{w(\tau)v(\tau+t)}=\int w(\tau)v(\tau+t)d\tau=0.
\label{eq:R_wv}
\end{align}
The response function $f(t)$ is obtained by deconvolution from the input-output cross-correlation.
The autocorrelation $R_{ww}(t)$ and the cross-correlation $R_{wy}(t)$ are,
\begin{subequations} 
\begin{align}
R_{ww}(t)&=\int w(\tau)w(\tau+t)d\tau,\\
R_{wy}(t)&=\int w(\tau)y(\tau+t)d\tau,\nonumber \\
&=\iint w(\tau)w(\tau')f(\tau-\tau'+t)d\tau d\tau',
\end{align}
\end{subequations} 
or in the frequency domain,
\begin{subequations} 
\begin{align}
S_{ww}(\omega)&=|W(\omega)|^2,\label{eq:auto_fft}\\
S_{wy}(\omega)&=|W(\omega)|^2F(\omega)\label{eq:cross_fft}.
\end{align}
\end{subequations} 
Therefore, the response function is obtained in the frequency domain by
\begin{align} 
F(\omega)=\frac{S_{wy}(\omega)}{S_{ww}(\omega)}.
\label{eq:F_omega}
\end{align}
The obtained response function $f(t)$ gives the optimal cancellation of the input signal, i.e., 
\begin{align}
\Braket{\left[y(t)-\int h(t-\tau)w(\tau)d\tau\right]^2}=\Braket{v^2(t)}+\Braket{\left\{\int \left[f(t-\tau)-h(t-\tau)\right]w(\tau)d\tau\right\}^2},
\end{align}
which is minimized when $h(t)=f(t)$.
Note that the cross terms vanish by using Eq.~\eqref{eq:R_wv}.

\subsection{Experimental response functions\label{subsec:response_function}}
If the QND gate is not working instantaneously, the QND gate transformations in the time domain are generally in the form of
\begin{subequations}
\begin{align}
\hat{x}_1^{\rm out}(t)&=\int f^{x}_{1\to1}(t-\tau)\hat{x}_1^{\rm in}(\tau)d\tau+(\text{other noise terms}),\\
\hat{x}_2^{\rm out}(t)&=\int f^{x}_{1\to2}(t-\tau)\hat{x}_1^{\rm in}(\tau)d\tau+\int f^{x}_{2\to2}(t-\tau)\hat{x}_2^{\rm in}(\tau)d\tau+(\text{other noise terms}),\\
\hat{p}_1^{\rm out}(t)&=\int f^{p}_{1\to1}(t-\tau)\hat{p}_1^{\rm in}(\tau)d\tau-\int f^{p}_{2\to1}(t-\tau)\hat{p}_2^{\rm in}(\tau)d\tau+(\text{other noise terms}),\\
\hat{p}_2^{\rm out}(t)&=\int f^{p}_{2\to2}(t-\tau)\hat{p}_2^{\rm in}(\tau)d\tau+(\text{other noise terms}).
\end{align}
\end{subequations}
We want to apply the theory in Sec.~\ref{subsec:general_theory} to this QND system.
For the estimation of the response functions, the random signals are used.
As an example, we consider the case where a random signal $\alpha(t)$ is added to the vacuum fluctuation $\hat{x}_1^{(0)}(t)$ as 
\begin{align}
\hat{x}^{\text{in}}_1(t)=\hat{x}_1^{(0)}(t)+\alpha(t),
\end{align}
and the other three quadratures are kept to vacuum levels.
In this case, in theory, by examining the transfer of the random signal $\alpha(t)$ to the two output quadratures $x^{\text{out}}_1(t)$ and $x^{\text{out}}_2(t)$, response functions $f_{1\to1}^{x}(t)$ and $f_{1\to2}^{x}(t)$ are obtained, respectively.
Note that the vacuum fluctuations, though they are white and random, cannot be used for the estimation of the response functions.
As discussed in Sec.~\ref{subsec:general_theory}, the important thing is that we know the input signal in order to obtain the cross correlation.
In reality, we cannot obtain the response functions with the procedures in Sec.~\ref{subsec:general_theory}.
The actual response functions obtained experimentally are $(f_{\text{in}\to k}\ast f_{k\to l}^{x, p}\ast f_{l\to\text{out}})(t)$, where $f_{\text{in}\to k}(t)$ is a response function of a conversion from an electronic signal to an optical signal, $f_{l\to\text{out}}(t)$ is a response function of a conversion from an optical signal to an electronic signal, and $\ast$ denotes a  convolution.

\begin{figure}[tb]
\centering
\includegraphics[width=17cm]{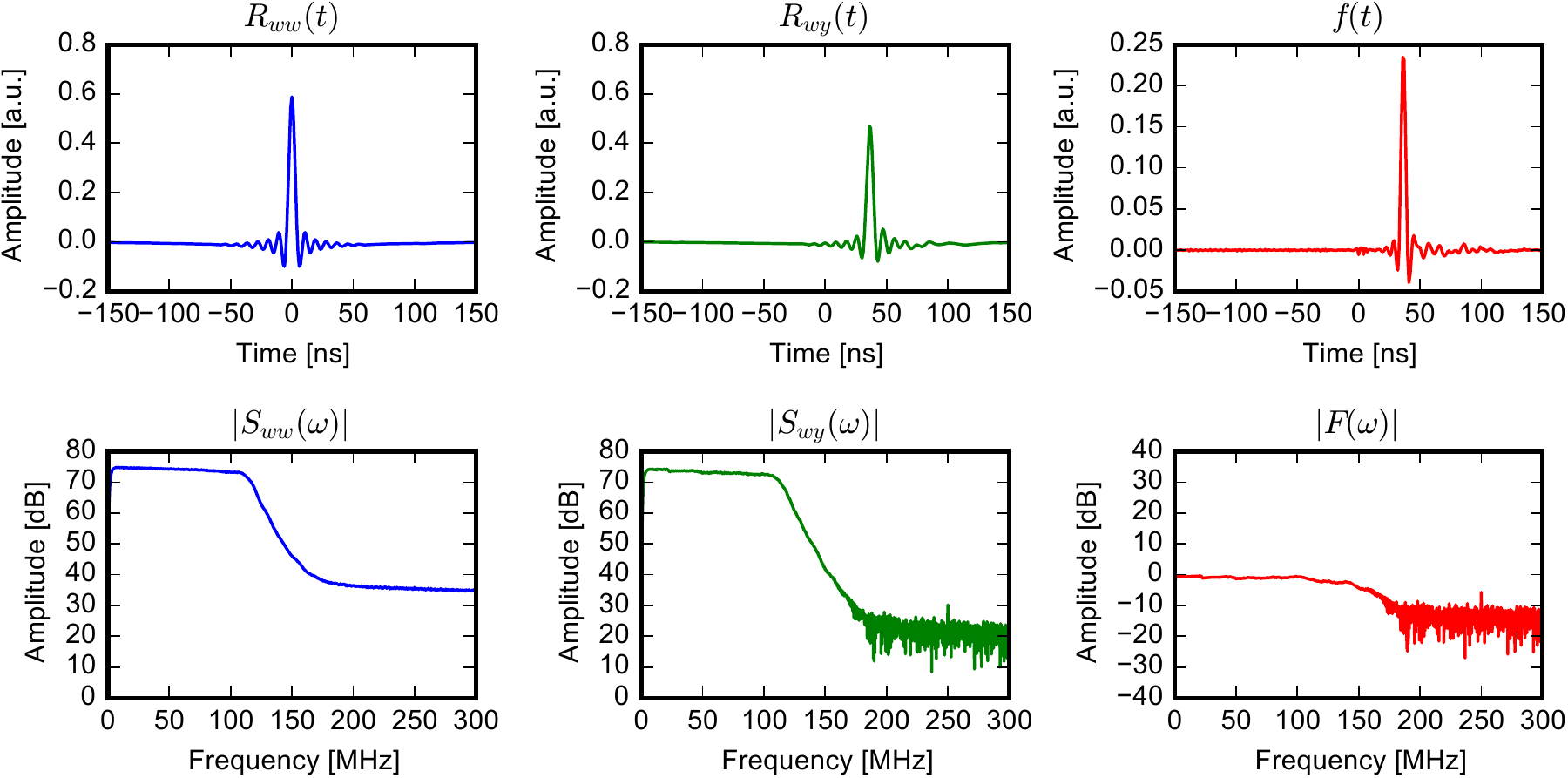}
\caption{\label{fig:response_INAX_X1}
Experimental input autocorrelation, input-output cross-correlation, and response function $(f_{\text{in}\to 1}\ast f_{1\to1}^x\ast f_{1\to\text{out}})(t)$ in the time domain (top panels) and in the frequency domain (bottom panels).
} 
\end{figure}

\begin{figure}[tb]
\centering
\includegraphics[width=10cm]{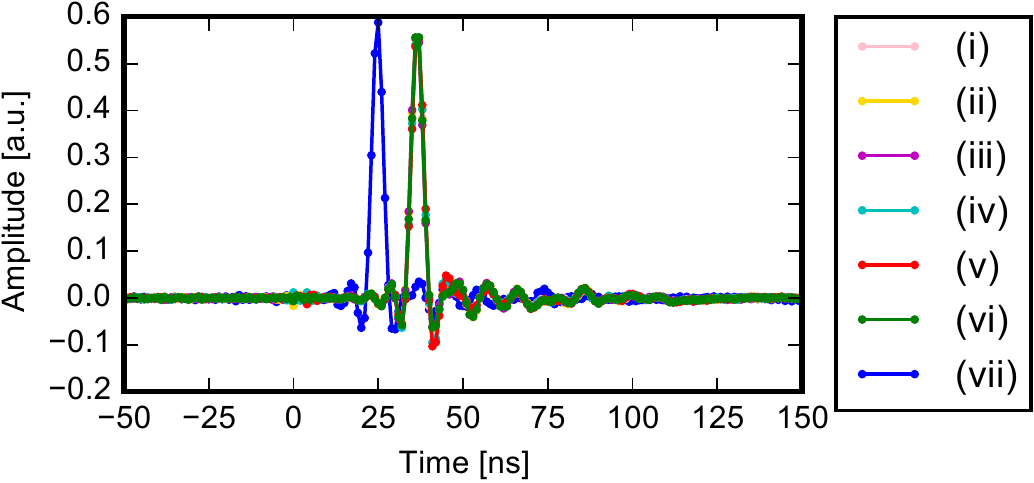}
\caption{\label{fig:transfer_func_exp}
Estimated response functions, used for the cancellation in Fig.~\ref{fig:coh_cancel4}.
(i) $(f_{\text{in}\to 1}\ast f_{1\to 1}^{x}\ast f_{1\to\text{out}})(t)$.
(ii) $(f_{\text{in}\to 1}\ast f_{1\to 2}^{x}\ast f_{2\to\text{out}})(t)$.
(iii) $(f_{\text{in}\to 2}\ast f_{2\to 2}^{x}\ast f_{2\to\text{out}})(t)$.
(iv) $(f_{\text{in}\to 1}\ast f_{1\to 1}^{p}\ast f_{1\to\text{out}})(t)$.
(v) $(f_{\text{in}\to 2}\ast f_{2\to 1}^{p}\ast f_{1\to\text{out}})(t)$.
(vi) $(f_{\text{in}\to 2}\ast f_{2\to 2}^{p}\ast f_{2\to\text{out}})(t)$.
(vii) $(f_{\text{in}\to 1}\ast f_{1\to\text{out}})(t)$.
}
\end{figure}

\begin{figure}[tb]
\centering
\includegraphics[width=17cm]{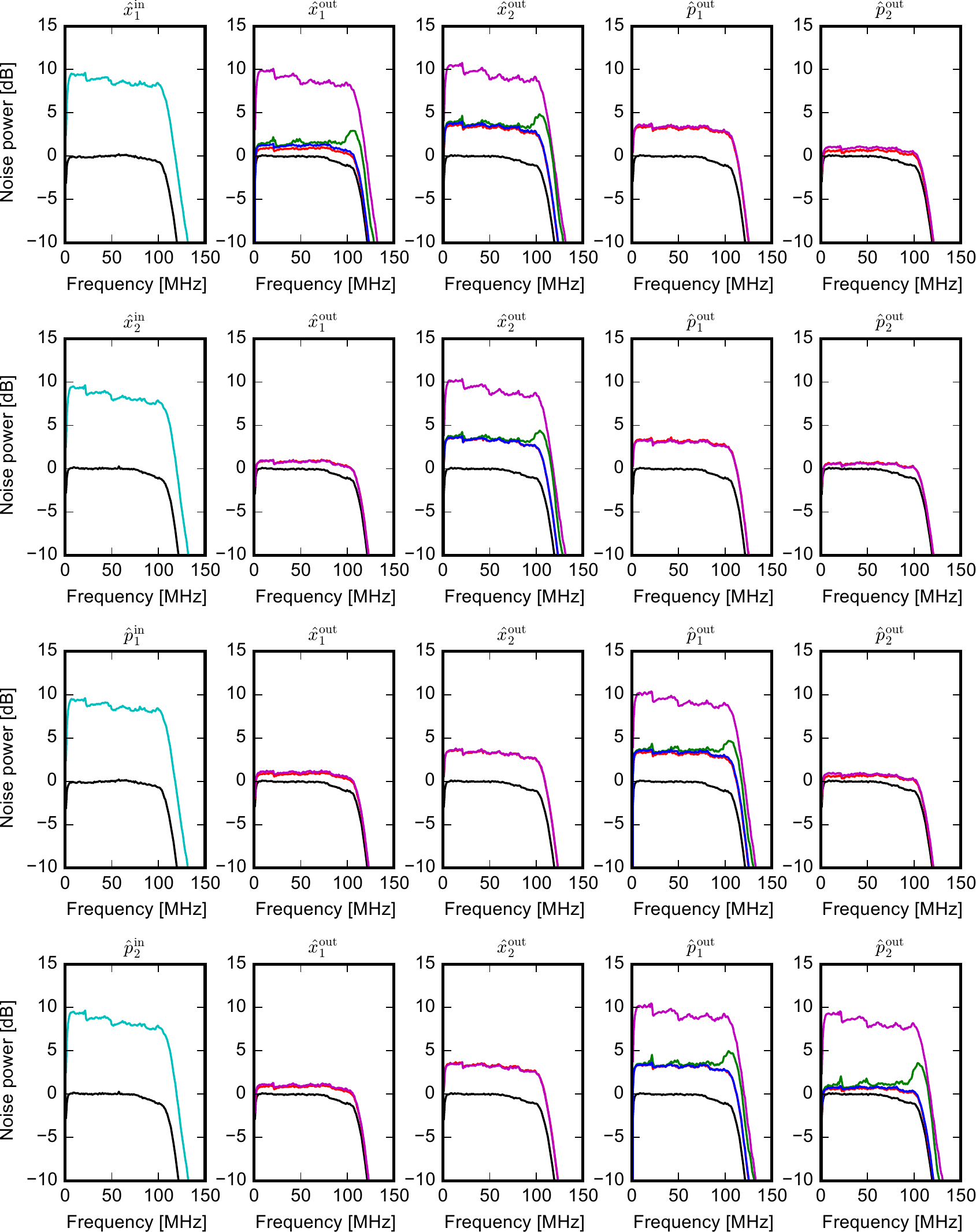}
\caption{\label{fig:coh_cancel4}
Power spectra of all the cases where a random signal is added to one of the input quadratures $\hat{x}_1^{\text{in}}$, $\hat{x}_2^{\text{in}}$, $\hat{p}_1^{\text{in}}$ and $\hat{p}_2^{\text{in}}$.
Black: shot noises.
Red: the QND outputs with vacuum-state inputs.
Cyan: optical random signal at the input.
Magenta: the QND outputs with the random signal input.
Green: cancellation of the random signal without the response functions.
Blue: cancellation of the random signal with the response functions.}
\end{figure}

\begin{figure}[tb]
\centering
\includegraphics[width=17cm]{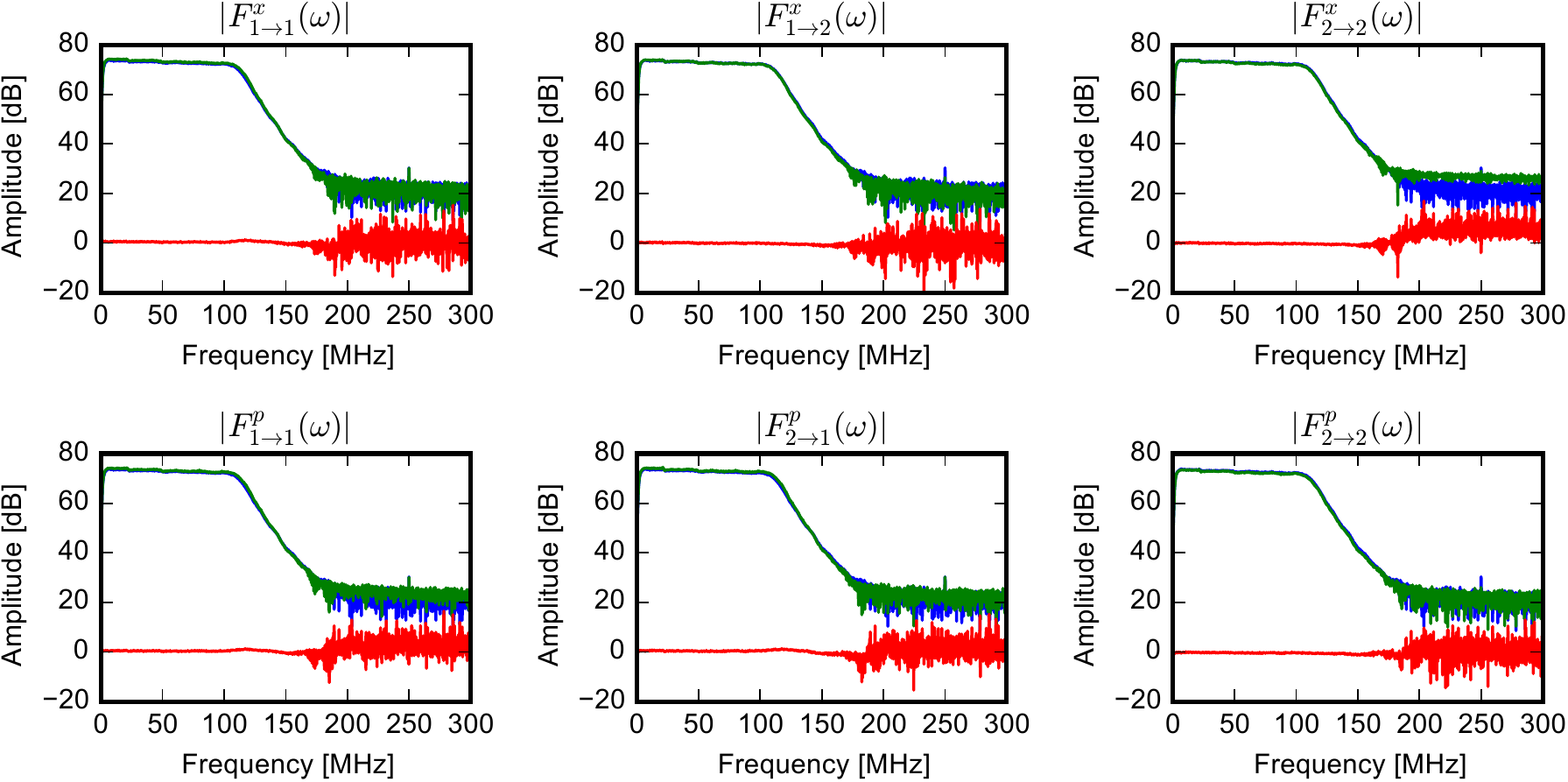}
\caption{\label{fig:tfQND}Response functions of the QND gate in the frequency domain. 
Blue: response function $|F_{\text{in}\to1}(\omega)F_{1\to\text{out}}(\omega)|$.
Green: response functions $|F_{\text{in}\to1}(\omega)F_{k\to l}^{x, p}(\omega)F_{1\to\text{out}}(\omega)|$.
Red: response functions $|F_{k\to l}^{x, p}(\omega)|$.
}
\end{figure}

\begin{table}[b]
\centering
\caption{Inner products between all the response functions (i)-(vii) in Fig.~\ref{fig:transfer_func_exp}.}
\label{tab:transfer_func}
\begin{tabular}{cccccccc} \hline \hline
Response function & (i) & (ii) & (iii) & (iv) & (v) & (vi) & (vii) \\ \hline \hline
(i) & 1 & 0.989 & 0.991 & 0.990 & 0.994 & 0.990 & 0.976 \\
(ii) & - & 1 & 0.989 & 0.990 & 0.988 & 0.995 & 0.978 \\
(iii) & - & - & 1 & 0.985 & 0.986 & 0.990 & 0.972 \\
(iv) & - & - & - & 1 & 0.992 & 0.991 & 0.982 \\
(v) & - & - & - & - & 1 & 0.989 & 0.982 \\
(vi) & - & - & - & - & - & 1 & 0.981 \\ 
(vii) & - & - & - & - & - & - & 1 \\
\hline \hline
\end{tabular}
\end{table}

As an example, we show the autocorrelation, the cross-correlation, and the obtained response function $(f_{\text{in}\to 1}\ast f_{1\to 1}^{x}\ast f_{1\to\text{out}})(t)$ in Fig.~\ref{fig:response_INAX_X1}.
All the other experimentally estimated response functions from the input electronic signals to output the electronic signals are shown as traces (i)--(vi) in Fig.~\ref{fig:transfer_func_exp}.
All the response functions have the same shape.
Note that, although we use LPFs and HPFs for output homodyne signals, the same filters are applied before the storage of the input signal as shown in Fig.~\ref{A-fig:setup}(b) in the main text and thus the effect of the filters are canceled in the response functions.
These response functions improve the cancellation in Fig.~\ref{A-fig:cancel} in the main text.
Figure \ref{fig:coh_cancel4} shows all the power spectra of cancellation with and without the response functions when the random signal is added to one of the four input quadratures $\hat{x}_1^{\rm in}$, $\hat{x}_2^{\rm in}$, $\hat{p}_1^{\rm in}$, and $\hat{p}_2^{\rm in}$.
Black, red, magenta, blue, and green traces are the spectra for the shot noises as references, the QND outputs with vacuum inputs, those with the random signal input, the cancellation with the response functions, and the cancellation without them, respectively.
The signals are perfectly cancelled when the response functions are used, which means that the evolution of the signals through the QND gate is completely predictable.  

However, we note that over-150-MHz components of  the response functions do not actually represent the response of the QND gate but are determined by other reasons.
For the frequencies higher than 150~MHz, the homodyne signals are highly attenuated by the LPF and thus electronic noises are dominant.
While these electronic noises have a negligible cross-correlation between channels, they contribute to the autocorrelation.
Even though we subtracted the background electronic noises obtained without the optical LOs, there were still some residual noises, by which the denominator becomes much larger than the numerator in Eq.~\eqref{eq:F_omega} over 150~MHz.
As a result, the response functions look as if they have a limited bandwidth of less than 150~MHz.
The dull shape of the response functions shown in Fig.~\ref{fig:transfer_func_exp} are because of these situations.

In order to estimate the response function of the QND gate itself $f_{k\to l}^{x, p}(t)$, we conducted the following experiment.
As references, we estimated the response functions $(f_{\text{in}\to 1}\ast f_{1\to\text{out}})(t)$ for conversion of electric signals to optical signals and vice versa without the QND gate, and obtained a trace (vii) in Fig.~\ref{fig:transfer_func_exp}.
Here, we assumed negligible differences among peripheral response functions $(f_{\text{in}\to k}\ast f_{l\to\text{out}})(t)$.
The trace (vii) has the same shape as those of the traces (i)-(vi) with a time difference of 11~ns.
This time difference simply represents the difference of the positions of the homodyne detectors, and does not directly represent the QND gate latency of about 13~ns corresponding to the optical path length of about 3.8~m.
The response functions of the QND gate are obtained by the deconvolution of the traces (i)-(vi) by the trace (vii), and the results in the frequency domain are shown in Fig.~\ref{fig:tfQND}.
The obtained spectra are flat for up to 100~MHz, and thus we conclude that the response functions of the QND gate are like a delta function in the considered time scale.
Inner products of all the traces (i)--(vii) in Fig~\ref{fig:transfer_func_exp} with the time shift of 11~ns are summarized in Tab.~\ref{tab:transfer_func}.

\section{\label{sec:QND_criteria}Transfer coefficients and conditional variances}
As discussed in Sec.~\ref{subsec:response_function}, the response of the QND gate is like a delta function in the considered time scale.
Therefore, we can apply the conventional QND criteria~\cite{A-Holland_PRA_1990} to the filtered quadrature values at each time, without considering a complicated mixing of quadratures at different times.
Here, we summarize the QND criteria, especially, the connections between the QND quantities and the signal-to-noise ratios (SNRs).

General linear conversions of a signal observable $\hat{A}_{\text{S}}$ and a probe observable $\hat{A}_{\text{P}}$ by a nonideal QND gate are
\begin{subequations}
\label{eq:general_QND}
\begin{align}
\hat{A}_{\rm S}^{\rm out}&=G_{\rm S,S}\hat{A}_{\rm S}^{\rm in}
+G_{\rm S,P}\hat{A}_{\rm P}^{\rm in}
+G_{\rm S,NC}\hat{N}_{\rm COM}+\hat{N}_{\rm S},\\
\hat{A}_{\rm P}^{\rm out}&=G_{\rm P,S}\hat{A}_{\rm S}^{\rm in}
+G_{\rm P,P}\hat{A}_{\rm P}^{\rm in}
+G_{\rm P,NC}\hat{N}_{\rm COM}+\hat{N}_{\rm P},
\end{align}
\end{subequations}
where $\hat{N}_{\rm COM}$ is a correlated component, and $\hat{N}_{\rm S}$ and $\hat{N}_{\rm P}$ are uncorrelated components, of excess noises of the gate. 
The success criteria of the QND measurements are \cite{A-Holland_PRA_1990},
\begin{align}
1&<T_{\rm S}+T_{\rm P},& V_{\rm S|P}&<1.\label{eq:QND_criteria}
\end{align} 
The transfer coefficients $T_{\rm S}$, $T_{\rm P}$ and the conditional variance $V_{\rm S|P}$ are defined as
\begin{subequations}
\begin{align}
T_{\rm S}&= C^2_{\hat{A}_{\rm S}^{\rm in}\hat{A}_{\rm S}^{\rm out}}=
\frac{|\braket{\hat{A}_{\rm S}^{\rm in}\hat{A}_{\rm S}^{\rm out}}-
\braket{\hat{A}_{\rm S}^{\rm in}}\braket{\hat{A}_{\rm S}^{\rm out}}|^2}
{V_{\hat{A}_{\rm S}^{\rm in}}V_{\hat{A}_{\rm S}^{\rm out}}},\label{eq:Ts}\\
T_{\rm P}&= C^2_{\hat{A}_{\rm S}^{\rm in}\hat{A}_{\rm P}^{\rm out}}=
\frac{|\braket{\hat{A}_{\rm S}^{\rm in}\hat{A}_{\rm P}^{\rm out}}-
\braket{\hat{A}_{\rm S}^{\rm in}}\braket{\hat{A}_{\rm P}^{\rm out}}|^2}
{V_{\hat{A}_{\rm S}^{\rm in}}V_{\hat{A}_{\rm P}^{\rm out}}},\label{eq:Tp}\\
V_{\rm S|P}&= V_{\hat{A}_{\rm S}^{\rm out}}
(1-C^2_{\hat{A}_{\rm S}^{\rm out}\hat{A}_{\rm P}^{\rm out}})\label{eq:Vsp}\notag \\
&=V_{\hat{A}_{\text{S}}^{\text{out}}}\left(1-\frac{V_{\hat{A}_{\text{S}}^{\text{out}}
\hat{A}_{\text{P}}^{\text{out}}}}{V_{\hat{A}_{\text{S}}^{\text{out}}}V_{\hat{A}_{\text{P}}^{\rm out}}}\right)
\notag\\
&=V_{\hat{A}_{\rm S}^{\rm out}}
\left(1-
\frac{|\braket{\hat{A}_{\rm S}^{\rm out}\hat{A}_{\rm P}^{\rm out}}-
\braket{\hat{A}_{\rm S}^{\rm out}}\braket{\hat{A}_{\rm P}^{\rm out}}|^2}
{V_{\hat{A}_{\rm S}^{\rm out}}V_{\hat{A}_{\rm P}^{\rm out}}}
\right),
\end{align}
\end{subequations}
where $V_{\hat{X}\hat{Y}}$, $V_{\hat{X}}$, and $C_{\hat{X}\hat{Y}}$ are a covariance, a variance, and a correlation, respectively,
\begin{subequations}
\begin{align}
V_{\hat{X}\hat{Y}}&=\braket{\hat{X}\hat{Y}}-\braket{\hat{X}}\braket{\hat{Y}},\\
V_{\hat{X}}&=V_{\hat{X}\hat{X}},\\
C_{\hat{X}\hat{Y}}&=\frac{V_{\hat{X}\hat{Y}}}{\sqrt{V_{\hat{X}}V_{\hat{Y}}}},
\end{align}
\end{subequations}
and the signal input state is assumed to be a coherent state, $V_{\hat{A}_{\text{S}}^{\text{in}}}=1$, i.e., the latter part of Eq.~\eqref{eq:QND_criteria} means that the signal observable is squeezed by the QND measurement.
Note that the transfer coefficients and the conditional variance are $T_{\rm S}=1$, $T_{\rm P}=G/(1+G)$, and $V_{\text{S}|\text{P}}=1/(1+G^2)$, for the ideal QND interaction, $\hat{A}_{\text{S}}^{\text{out}}=\hat{A}_{\text{S}}^{\text{in}}$, $\hat{A}_{\text{P}}^{\text{out}}=G\hat{A}_{\text{S}}^{\text{in}}+\hat{A}_{\text{P}}^{\text{in}}$, with a coherent-state probe input $V_{\hat{A}_{\text{P}}^{\text{in}}}=1$.
The excess noises of the gate decrease the transfer coefficients and increase the conditional variance.
With the general linear conversions in Eq.~\eqref{eq:general_QND}, the transfer coefficients are
\begin{subequations}  
\begin{align}
T_{\rm S}&=\frac{G_{\rm{S,S}}^2V_{\hat{A}_{\rm S}^{\rm in}}}
{G_{\rm{S,S}}^2V_{\hat{A}_{\rm S}^{\rm in}}
+G_{\rm{S,P}}^2V_{\hat{A}_{\rm P}^{\rm in}}
+G_{\rm{S,NC}}^2V_{\hat{N}_{\rm COM}}
+V_{\hat{N}_{\rm S}}
},
\\
T_{\rm P}&=\frac{G_{\rm{P,S}}^2V_{\hat{A}_{\rm S}^{\rm in}}}
{G_{\rm{P,S}}^2V_{\hat{A}_{\rm S}^{\rm in}}
+G_{\rm{P,P}}^2V_{\hat{A}_{\rm P}^{\rm in}}
+G_{\rm{P,NC}}^2V_{\hat{N}_{\rm COM}}
+V_{\hat{N}_{\rm P}}
},
\end{align}
\label{eq:Ts_Tp}
\end{subequations}
and the conditional variance is discussed later.
 
The transfer coefficients $T_{\rm S}$ and $T_{\rm P}$ are experimentally obtained by examining the transfer of the SNRs.
For this purpose, we add a signal $\alpha$ to the signal input $\hat{A}_{\text{S}}^{\text{in}}=\delta\hat{A}_{\text{S}}^{\text{in}}+\alpha$, where $\delta\hat{A}_{\text{S}}^{\text{in}}$ is a vacuum noise fluctuation $\braket{\delta\hat{A}_{\text{S}}^{\text{in}}}=0$, $V_{\delta\hat{A}_{\text{S}}^{\text{in}}}=1$, and the power is compared with that of the case without the input signal $\hat{A}_{\text{S}}^{\text{in}}=\delta\hat{A}_{\text{S}}^{\text{in}}$.
The SNR at the signal input is
\begin{align}
\label{eq:SNR_in}
{\rm SNR}_{\text{S}}^{\rm in}
=\frac{\alpha^2}{V_{\hat{A}_{\rm S}^\text{in}}}=
\frac{
\braket{(\delta\hat{A}_{\rm S}^{\rm in}+\alpha)^2}
-\braket{(\delta\hat{A}_{\rm S}^{\rm in})^2}
}
{
\braket{(\delta\hat{A}_{\rm S}^{\rm in})^2}
},
\end{align}
and thus obtained experimentally from the powers of the two cases $\braket{(\delta\hat{A}_{\rm S}^{\rm in}+\alpha)^2}$ and $\braket{(\delta\hat{A}_{\rm S}^{\rm in})^2}$.
On the other hand, the output signal and probe observables become $\hat{A}_{\text{S}}^{\text{out}}=\delta\hat{A}_{\text{S}}^{\text{out}}+G_{\text{S,S}}\alpha$ and 
$\hat{A}_{\text{P}}^{\text{out}}=\delta\hat{A}_{\text{P}}^{\text{out}}+G_{\text{P,S}}\alpha$, where $\delta\hat{A}_{\text{S}}^{\text{out}}$ and $\delta\hat{A}_{\text{P}}^{\text{out}}$ are noise fluctuations without the input signal $\alpha$.
We assume $\braket{\delta\hat{A}_{\text{S}}^{\text{out}}}=\braket{\delta\hat{A}_{\text{P}}^{\text{out}}}=0$ without loss of generality.
The SNRs at the signal and probe outputs are,
\begin{subequations}
\label{eq:SNR_out}
\begin{align}
{\rm SNR}_{\rm S}^{\rm out}
&=
\frac{G_{\text{S,S}}^2\alpha^2}{V_{\hat{A}_{\text{S}}^{\text{out}}}}
=
\frac{
\braket{(\delta\hat{A}_{\rm S}^{\rm out}+G_{\text{S,S}}\alpha)^2}
-\braket{(\delta\hat{A}_{\rm S}^{\rm out})^2}
}
{
\braket{(\delta\hat{A}_{\rm S}^{\rm out})^2}
}
\notag\\
&=
\frac{
G_{\rm S,S}^2\alpha^2
}
{
G_{\rm{S,S}}^2V_{\hat{A}_{\rm S}^{\rm in}}
+G_{\rm{S,P}}^2V_{\hat{A}_{\rm P}^{\rm in}}
+G_{\rm{S,NC}}^2V_{\hat{N}_{\rm COM}}
+V_{\hat{N}_{\rm S}}
},\label{SNR_S_out}
\\
{\rm SNR}_{\rm P}^{\rm out}
&=
\frac{G_{\text{P,S}}^2\alpha^2}{V_{\hat{A}_{\text{P}}^{\text{out}}}}
=
\frac{
\braket{(\delta\hat{A}_{\rm P}^{\rm out}+G_{\text{P,S}}\alpha)^2}
-\braket{(\delta\hat{A}_{\rm P}^{\rm out})^2}
}
{
\braket{(\delta\hat{A}_{\rm P}^{\rm out})^2}
}
\notag\\
&=
\frac{
G_{\rm P,S}^2\alpha^2
}
{
G_{\rm{P,S}}^2V_{\hat{A}_{\rm S}^{\rm in}}
+G_{\rm{P,P}}^2V_{\hat{A}_{\rm P}^{\rm in}}
+G_{\rm{P,NC}}^2V_{\hat{N}_{\rm COM}}
+V_{\hat{N}_{\rm S}}
}.
\end{align}
\end{subequations}
and thus obtained experimentally from the powers of the two cases $\braket{(\delta\hat{A}_{\rm S}^{\rm out}+G_{\text{S,S}}\alpha)^2}$, 
$\braket{(\delta\hat{A}_{\rm P}^{\rm out}+G_{\text{P,S}}\alpha)^2}$, and
$\braket{(\delta\hat{A}_{\rm S}^{\rm out})^2}$, 
$\braket{(\delta\hat{A}_{\rm P}^{\rm out})^2}$.
By using Eqs.~\eqref{eq:Ts_Tp}--\eqref{eq:SNR_out}, we obtain, 
\begin{align}
T_{\rm S}&=\frac{{\rm SNR}_{\rm S}^{\rm out}}{{\rm SNR}_{\rm S}^{\rm in}},
&
T_{\rm P}&=\frac{{\rm SNR}_{\rm P}^{\rm out}}{{\rm SNR}_{\rm S}^{\rm in}}.
\end{align}
Therefore, $T_{\text{S}}$ and $T_{\text{P}}$ represent the degradation of the SNR when the signal input $\alpha$ is transferred to the signal and probe outputs, respectively.

The conditional variance $V_{\rm S|P}$ corresponds to the minimum variance of $\hat{A}_{\rm S}^{\rm out}-g\hat{A}_{\rm P}^{\rm out}$ where the subtraction gain $g$ is an optimization parameter.
The variance of $\hat{A}_{\rm S}^{\rm out}-g\hat{A}_{\rm P}^{\rm out}$ is a quadratic polynomial in $g$,
\begin{align}
V_{\hat{A}_{\rm S}^{\rm out}-g\hat{A}_{\rm P}^{\rm out}}=
&\braket{(\delta\hat{A}_{\rm S}^{\rm out}-g\delta\hat{A}_{\rm P}^{\rm out})^2}
\notag\\
=&
V_{\hat{A}_{\rm P}^{\rm out}}g^2-
2V_{\hat{A}_{\rm S}^{\rm out}\hat{A}_{\rm P}^{\rm out}}g
+V_{\hat{A}_{\rm S}^{\rm out}}\notag\\
=&
V_{\hat{A}_{\rm P}^{\rm out}}\left(
g-\frac{V_{\hat{A}_{\rm S}^{\rm out}\hat{A}_{\rm P}^{\rm out}}}
{V_{\hat{A}_{\rm P}^{\rm out}}}
\right)^2+V_{\text{S}|\text{P}},
\end{align}
which is minimized at $g=V_{\hat{A}_{\rm S}^{\rm out}\hat{A}_{\rm P}^{\rm out}}
/V_{\hat{A}_{\rm P}^{\rm out}}$.

\begin{figure}[tb]
\centering
\includegraphics[width=10cm]{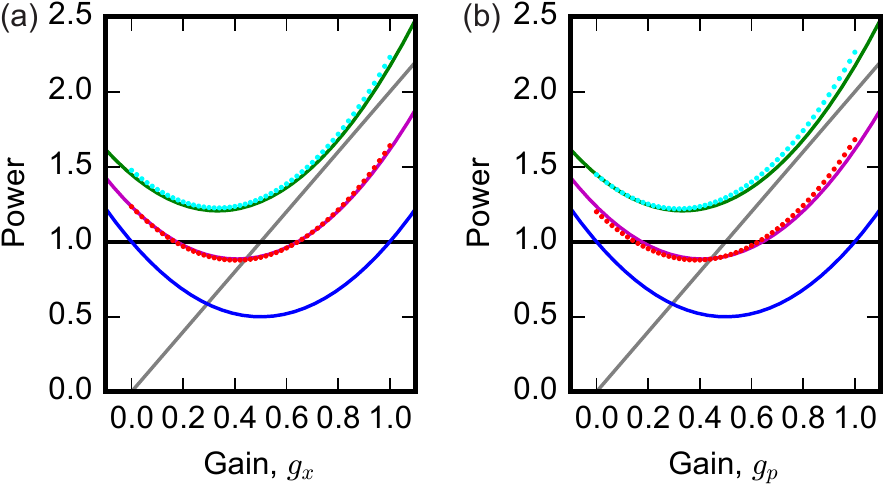}
\caption{\label{fig:cond_var}
Variances of (a) $\hat{x}_1^{\rm out}-g_x\hat{x}_2^{\rm out}$ and (b) $g_p\hat{p}_1^{\rm out}+\hat{p}_2^{\rm out}$.
Red markers: experimental results when ancillary squeezed vacua are used. 
Cyan markers: experimental results when ancillary squeezed vacua are not used. 
Magenta curves: theoretical variance when ancillary squeezed vacua with $-2.8$ dB of squeezing are used. 
Green curves: theoretical variance when ancillary squeezed vacua are not used. 
Blue curves: theoretical variance for the ideal QND interaction with $G=1$. 
Gray lines: entangled criterion.}
\end{figure}

The experimental values are summarized in Tab.~\ref{tab:QND_criteria}.
The variances $\braket{(\hat{x}_1^{\rm out}-g_x\hat{x}_2^{\rm out})^2}$ and $\braket{(g_p\hat{p}_1^{\rm out}+\hat{p}_2^{\rm out})^2}$ for various subtraction and addition gains are plotted in Fig.~\ref{fig:cond_var}. 
\begin{table}[tb]
\centering
\caption{Verification of transfer coefficients in the QND gate. The coherent state amplitude is injected either to $x_{1}^{\rm in}$ or $p_{2}^{\rm in}$.}
\begin{tabular}{ccc} \hline \hline
Coherent state input & $x_{1}^{\rm in}$ & $p_{2}^{\rm in}$ \\ \hline \hline
$T_{\rm S}$ & 0.86$\pm$0.06 & 0.85$\pm$0.06 \\
$T_{\rm P}$ & 0.51$\pm$0.04 & 0.52$\pm$0.04 \\
$T_{\rm S}+T_{\rm P}$ & 1.37$\pm$0.10 & 1.37$\pm$0.10 \\ \hline
$V_{\rm S|P}$ & 0.88$\pm$0.03 & 0.88$\pm$0.03 \\ \hline \hline
\end{tabular}
\label{tab:QND_criteria}
\end{table}

\section{Quantum Entanglement}
The sub-shot-noise conditional variances $V_{\rm S|P}<1$ in both of the $\hat{x}$ and $\hat{p}$ quadratures are not a sufficient condition for entanglement.
A sufficient condition based on the Duan-Simon criterion is~\cite{A-Yoshikawa_PRL_2008, A-Duan_PRL_2000, A-Simon_PRL_2000},
\begin{align}
\exists g,\quad\braket{(\hat{x}_1^{\text{out}}-g\hat{x}_2^{\text{out}})^2}+\braket{(g\hat{p}_1^{\text{out}}+g\hat{p}_2^{\text{out}})^2}<4|g|.
\end{align}
In Fig.~\ref{fig:cond_var}, there are gray lines $\braket{(\hat{x}_1^{\text{out}}-g\hat{x}_2^{\text{out}})^2}=2|g|$ and $\braket{(g\hat{p}_1^{\text{out}}+\hat{p}_2^{\text{out}})^2}=2|g|$, and there is a region of $g$ where red markers are below the gray lines in both quadratures.
Therefore, the two output modes are entangled for coherent-state inputs.

\end{document}

%% file: author_list.tex

\author{Yu Shiozawa} \affiliation{Department of Applied Physics, School of Engineering, The University of Tokyo, 7-3-1 Hongo, Bunkyo-ku, Tokyo 113-8656, Japan}

\author{Jun-ichi Yoshikawa} \affiliation{Department of Applied Physics, School of Engineering, The University of Tokyo, 7-3-1 Hongo, Bunkyo-ku, Tokyo 113-8656, Japan}
\affiliation{Quantum-Phase Electronics Center, School of Enginerring, The University of Tokyo, 7-3-1 Hongo, Bunkyo-ku, Tokyo 113-8656, Japan}

\author{Shota Yokoyama} \affiliation{Department of Applied Physics, School of Engineering, The University of Tokyo, 7-3-1 Hongo, Bunkyo-ku, Tokyo 113-8656, Japan}
\affiliation{Centre for Quantum Computation and Communication Technology, School of Engineering and Information Technology, University of New South Wales Canberra, ACT 2600, Australia}

\author{Toshiyuki Kaji} \affiliation{Department of Applied Physics, School of Engineering, The University of Tokyo, 7-3-1 Hongo, Bunkyo-ku, Tokyo 113-8656, Japan}

\author{Kenzo Makino} \affiliation{Department of Applied Physics, School of Engineering, The University of Tokyo, 7-3-1 Hongo, Bunkyo-ku, Tokyo 113-8656, Japan}

\author{Takahiro Serikawa} \affiliation{Department of Applied Physics, School of Engineering, The University of Tokyo, 7-3-1 Hongo, Bunkyo-ku, Tokyo 113-8656, Japan}

\author{Ryosuke Nakamura} \affiliation{Department of Applied Physics, School of Engineering, The University of Tokyo, 7-3-1 Hongo, Bunkyo-ku, Tokyo 113-8656, Japan}

\author{Shigenari Suzuki} \affiliation{Department of Applied Physics, School of Engineering, The University of Tokyo, 7-3-1 Hongo, Bunkyo-ku, Tokyo 113-8656, Japan}

\author{Shota Yamazaki} \affiliation{Department of Applied Physics, School of Engineering, The University of Tokyo, 7-3-1 Hongo, Bunkyo-ku, Tokyo 113-8656, Japan}

\author{Warit Asavanant} \affiliation{Department of Applied Physics, School of Engineering, The University of Tokyo, 7-3-1 Hongo, Bunkyo-ku, Tokyo 113-8656, Japan}

\author{Shuntaro Takeda} \affiliation{Department of Applied Physics, School of Engineering, The University of Tokyo, 7-3-1 Hongo, Bunkyo-ku, Tokyo 113-8656, Japan}
\affiliation{JST, PRESTO, 4-1-8 Honcho, Kawaguchi, Saitama, 332-0012, Japan}

\author{Peter van Loock} \affiliation{Institute of Physics, Johannes Gutenberg-Universit\"at Mainz, Staudingerweg 7, 55099 Mainz, Germany}

\author{Akira Furusawa} \email{akiraf@ap.t.u-tokyo.ac.jp}\affiliation{Department of Applied Physics, School of Engineering, The University of Tokyo, 7-3-1 Hongo, Bunkyo-ku, Tokyo 113-8656, Japan}

\vskip 0.25cm